\documentclass[fleqn,usenatbib]{mnras}

\usepackage{newtxtext,newtxmath}

\usepackage[T1]{fontenc}

\DeclareRobustCommand{\VAN}[3]{#2}
\let\VANthebibliography\thebibliography
\def\thebibliography{\DeclareRobustCommand{\VAN}[3]{##3}\VANthebibliography}

\usepackage{graphicx}	
\usepackage{amsmath}	
\usepackage{hyperref}
\hypersetup{colorlinks=true, linkcolor=red, urlcolor=black, citecolor=blue}
\usepackage{graphicx}
\usepackage{siunitx}
\DeclareSIUnit\angstrom{\text {Å}}
\usepackage{booktabs}
\usepackage{bm}
\defcitealias{Mestric2022}{M22}
\newcommand*\diff{\mathop{}\!\mathrm{d}}

\title[UV-continuum $\beta$ slopes of individual lensed clumps]{UV-continuum $\beta$ slopes of individual $z \sim 2-6$ clumps and their evolution}
\author[A. Bolamperti et al.]{A.~Bolamperti$^{1,2,3}$\thanks{E-mail: andrea.bolamperti@phd.unipd.it},
A.~Zanella$^{2}$,
U.~Me\v{s}tri\'{c}$^{4,5}$,
E.~Vanzella$^{5}$,
M.~Castellano$^{6}$,
P.~Bergamini$^{4,5}$,
F.~Calura$^{5}$,
\newauthor
C.~Grillo$^{4}$,
M.~Meneghetti$^{5}$,
A.~Mercurio$^{7,8}$
P.~Rosati$^{9,5}$,
T.~Devereaux$^{1}$,
E.~Iani$^{10}$,
J.~Vernet$^{3}$.
\\ \\
$^{1}$Dipartimento di Fisica e Astronomia, Università degli Studi di Padova, Vicolo dell'Osservatorio 3, I-35122 Padova, Italy\\
$^{2}$Istituto Nazionale di Astrofisica (INAF), Osservatorio di Padova, Vicolo dell'Osservatorio 5, I-35122 Padova, Italy\\
$^{3}$European Southern Observatory, Karl-Schwarzschild-Strasse 2, D-85748 Garching bei M\"unchen, Germany\\
$^{4}$Dipartimento di Fisica, Università degli Studi di Milano, Via Celoria 16, I-20133 Milano, Italy \\
$^{5}$INAF -- OAS, Osservatorio di Astrofisica e Scienza dello Spazio di Bologna, via Gobetti 93/3, I-40129 Bologna, Italy\\
$^{6}$INAF Osservatorio Astronomico di Roma, Via Frascati 33, 00078 Monteporzio Catone, Rome, Italy \\
$^{7}$Department of Physics of the University of Salerno - Via Giovanni Paolo II, 132, 84084, Fisciano (SA), Italy\\
$^{8}$INAF-Astronomical Observatory of Capodimonte, Salita Moiariello 16, I-80131 Napoli, Italy \\
$^{9}$Dipartimento di Fisica e Scienze della Terra, Università degli Studi di Ferrara, Via Saragat 1, I-44122 Ferrara, Italy\\
$^{10}$Kapteyn Astronomical Institute, University of Groningen, 9700AV Groningen, The Netherlands
}
\date{Accepted XXX. Received YYY; in original form ZZZ}
\pubyear{2023}

\begin{document}
\label{firstpage}
\pagerange{\pageref{firstpage}--\pageref{lastpage}}
\maketitle

\begin{abstract}
We study the ultraviolet (UV) continuum $\beta$ slope of a sample of 166 clumps, individual star-forming regions observed in high redshift galaxies. They are hosted by 67 galaxies with redshift between 2 and 6.2, strongly lensed by the Hubble Frontier Fields cluster of galaxies MACS~J0416.1$-$2403. The $\beta$ slope is sensitive to a variety of physical properties, such as the metallicity, the age of the stellar population, the dust attenuation throughout the galaxy, the stellar initial mass function (IMF), and the star-formation history (SFH). The aim of this study is to compare the $\beta$ values of individual clumps with those measured on the entire galaxy, to investigate possible physical differences between these regions and their hosts. We found a median value of $\beta \sim -2.4$, lower than that of integrated galaxies. This result confirms that clumps are sites of intense star formation, populated by young, massive stars, whose spectrum strongly emits in the UV. This is also consistent with the assumption that the dust extinction at the location of the clumps is lower than the average extinction of the galaxy, or that clumps have a different IMF or SFH. We made use of the correlations, discovered for high-redshift galaxies, of the $\beta$ value with those of redshift and UV magnitude, $M_{UV}$, finding that clumps follow the same relations, extended to much fainter magnitudes ($M_{UV}<-13$).
We also find evidence of eight clumps with extremely blue ($\beta \lesssim -2.7$) slopes, which could be the signpost of low-metallicity stars and constrain the emissivity of ionizing photons at high redshift.

\end{abstract}

\begin{keywords}
galaxies: high-redshift -- galaxies: evolution 
\end{keywords}



\section{Introduction}

In the last years, the characterization of observed \citep[e.g.,][]{Madau2014, ForsterSchreiber2009, Dunlop2012, Murata2014, Zanella2015, Bouwens2015, Stott2016, Guo2018, Oesch2018, Romano2021, Sommovigo2021, Vanzella2022} and simulated \citep[e.g.,][]{Bournaud2014, Tamburello2015, Buck2017, Lovell2021, Zanella2021, Pallottini2022, Vizgan2022, Calura2022} galaxies at high redshift represented a key step in the study of galaxy evolution in the early Universe. 
Recently, thanks to the first data from the James Webb Space Telescope (\textit{JWST}), some studies have reached new and unexplored epochs, discovering galaxies up to $z\sim 10$--$17$ \citep{Atek2022,Castellano2022, Donnan2022, Finkelstein2022, Harikane2022}, just approximately $200$--$500$ Myr after the Big Bang in the standard cosmological model.

Galaxies at high-$z$ show a morphology different from local ones. They consist of turbulent disky structures, i.e., marginally-stable rotating discs with a significant contribution of random motions to the dynamical support of the system \citep{Elmegreen2007, Glazebrook2013, Guo2015, Ferreira22}. These galaxies are dominated by bright blue knots, dubbed \textit{clumps}, visible also in the latest \textit{JWST} images \citep[e.g.,~][]{Treu2022b}. Observations showed that clumps have stellar mass values between $10^7$ and $10^9$\,M$_\odot$ \citep{ForsterSchreiber2011a, Guo2012, Soto2017}, star-formation rate values between $0.1$ and $10$ M$_\odot$\,yr$^{-1}$ \citep{Guo2012, Soto2017} and are star-forming regions, i.e., with a specific star-formation rate considerably larger than their host galaxy \citep{Bournaud2015, Zanella2015, Zanella2019}. Clumps in field galaxies are unresolved (FWHM $< 1$ kpc) \citep{Elmegreen2007, ForsterSchreiber2011b, Genzel2011}, and it has been shown, through simulations and observations with different resolutions \citep{Oklopic2017, Behrendt2016, Behrendt2019, Tamburello2017, Faure2021, Mestric2022, Claeyssens2023}, that the detectability and the measured sizes of clumps depend on the observational resolution. Thanks to gravitational lensing, we can observe clump sizes of the order of hundreds of pc \citep{Livermore2015, Rigby2017, Cava2018, Dessauges-Zavadsky2017, Dessauges-Zavadsky2019}, down to a few tens of pc in extremely high-magnification regimes \citep{Johnson2017, Vanzella2019, Vanzella2020a, Calura2021, Vanzella2022, Mestric2022, Messa2022, Claeyssens2023}. Although gravitational lensing remains the best opportunity to explore compact and faint structures, a proper correction of the observed properties to infer the intrinsic ones, i.e., the \textit{delensed} ones, requires the most accurate lensing models of cluster of galaxies developed nowadays, with the largest number of constraints \citep[e.g.,][]{Caminha2017, Bonamigo2018, Bergamini2021, Bergamini2022}. Their accuracy critically depends on the number of spectroscopically-confirmed multiple images, to minimize the number of misidentified systems and the degeneracy between the observer-deflector-source relative distances and the total mass distribution of the deflector \citep{Johnson2014, Grillo2015, Caminha2019, Bolamperti2023}. 

In this framework, the investigation of the physical properties of galactic substructures and individual clumps hosted by high-$z$ galaxies down to the smallest scales gives unique hints to study galaxy evolution. One of the most important quantities exploited to characterize young stellar populations is the UV-continuum slope. This can be estimated directly from multi-band rest-frame UV measurements, or from spectral energy distribution (SED) fitting. The UV-continuum slope is commonly referred to as ``$\beta$ slope'', because it has been shown that the UV part of the spectrum can reasonably be fitted with a simple power-law relation, i.e., $f_\lambda \propto \lambda^\beta$ \citep{Calzetti1994}. Despite some degeneracy with the metallicity of a galaxy \citep[e.g.][]{Castellano2014, Calabro2021} and with its star formation history \citep[SFH, ][]{Bouwens2016}, the value of $\beta$ can give important insights on the stellar population, in particular about the presence of young stars. It is also a common indicator of dust attenuation \citep{Calzetti1994}, allowing the study of the evolution and build-up of dust in galaxies, from $z \sim 2$ to 10 \citep{Reddy2018}. The intrinsic slope is the combination of different factors defining the stellar population in the galaxy, such as the total amount and composition of dust grains, the metallicity, the stellar initial mass function (IMF), and the star formation history (SFH). It is not possible to constrain all these physical parameters separately, but a robust measurement of the $\beta$ slope is necessary to characterize the global properties of galaxies. As a reference, a galaxy with a dust-free stellar population with solar metallicity and constant star formation rate has a value of $\beta \simeq -2.2$. Several studies, based on HST data up to $z \sim 8$, show average $\beta$ values $\lesssim -2$, a sign of young and metal-poor stellar populations \citep[e.g.,][]{Dunlop2013, Finkelstein2012, Bouwens2014, castellano2023ion}. In fact, the bluest $z = 2-4$ \citep{Castellano2012, McLure2018} and local \citep{Calzetti1994, Vazquez2004} galaxies usually show slopes between $-2.5$ and $-2$. 
Great efforts have been dedicated to the search for the bluest galaxies at high-$z$. Within the photometric uncertainties increasing for the bluest slopes, the discovery of robust $\beta \simeq -3$ candidates,  indicating stellar populations formed from pristine gas with a large ionizing photon escape fraction, would provide important insights into the composition and characteristics of the most distant galaxies.
Many models predict that, in principle, slopes of $-3$ would be produced by extremely low metallicity and young stellar populations \citep{Raiter2010, Bouwens2010, Topping2022}. However, we do not expect to observe them because of the presence of nebular continuum emission from the ionized gas around young stars, that reddens the slopes up to $\Delta \beta \sim 0.5$ \citep{Raiter2010, Bouwens2010, Trussler2022}, even if this effect may be mitigated if the ionizing radiation leaks directly into the intergalactic medium (IGM). Despite this, some studies presented $\beta \lesssim -3$ slopes, not always reproducible with stellar population models \citep{Bouwens2010, Labbe2010, Ono2010, Jiang2020, Topping2022}.
Particular attention has been dedicated to the identification of reionization-era galaxies with very blue UV slopes, whose young, low-metallicity stellar populations and large escape fraction of ionizing photons into the IGM can be interpreted as the presence of zero metallicity, massive PopIII stars \citep{Wise2012, Dayal2018}. 

So far, no systematic studies on the $\beta$ slopes of individual star-forming clumps have been done. The building of a significant sample of individual star-forming clumps over a broad redshift range would be essential to characterize their physical properties and the interplay with their host galaxies, which is also a key ingredient in high-resolution hydrodynamical simulations. 

In this paper we investigate the UV-continuum slope of individual clumps between redshift of approximately 2 and 6, and discuss several factors that can affect these measurements. This paper is organized as follows. In Section 2, we present the ancillary data, discuss the sample selection, and summarize the algorithm used to identify individual clumps. In Section 3, we detail the process we used to extract photometric measurements and to measure the $\beta$ slope. In Section 4, we do the same with spectroscopic data, and compare the results with the photometric slopes. In Section 5, we compare our individual clump results with those for galaxies from the literature, and discuss trends of $\beta$ with magnitude and redshift. At the end of the section, we also discuss some cases that exhibit an extremely blue slope ($\beta \lesssim -2.7$), analyzing the physical scenarios that can explain them. Finally, in Section 6 we summarize the results and draw conclusions, discussing the caveats of this study.

Throughout this work, we assume a flat cosmology with $H_0 = 70 \, \si{km.s^{-1}.Mpc^{-1}}$, $\Omega_m = 0.3$ and $\Omega_\Lambda = 0.7$. Unless otherwise specified, all magnitudes are given in the AB system.

\section{Sample selection and ancillary data}
\label{sec:data}

\subsection{\textit{HST} data}
MACS~J0416.1$-$2403 (hereafter, MACS~J0416), (RA, dec) $=$ (04:16:08.9, $-24$:04:28.7) at $z=0.396$, is one of the galaxy clusters which act as gravitational lenses with the largest number of observed multiple images \citep{Zitrin2013}, likely because of its highly elongated and irregular structure. It has been included in the Hubble Frontier Field (HFF) program and thus observed in seven \textit{HST} filters. 
We make use of the deep, multi-wavelength observations in the MACS~J0416 field, which are publicly available \citep{Lotz2017, Koekemoer2014}, and of {\tt ASTRODEEP} PSF matched images in the \textit{HST}/ACS F435W, F606W, F814W, and \textit{HST}/WFC3 F105W, F125W, F140W, and F160W bands \citep{Merlin2016a, Castellano2016}. In the following, we will refer to \textit{HST}/ACS F435W, F606W, F814W, and \textit{HST}/WFC3 F105W, F125W, F140W, and F160W bands as, respectively, \textit{B435, V606, I814, Y105, J125, JH140, H160}. \cite{Merlin2016a} complemented the HFF data with imaging from the CLASH survey \citep[PI: M. Postman,][]{Postman2012} and program 13386 (PI: S. Rodney), and use the final reduced and calibrated v1.0 mosaics released by the Space Telescope Science Institute (STScI), drizzled at $0.06''$ pixel-scale. The \textit{H160} image plays a key role in this framework, because it has the worst PSF FWHM ($0.20''$), which increases with increasing wavelength (e.g., from $0.11''$ in the \textit{B435} to $0.19''$ in the \textit{JH140} band). All the images in the remaining six filters have been PSF-matched to the \textit{H160} one, with a convolution kernel that was obtained from the ratio of the PSFs of the respective pair of images in the Fourier space. Furthermore, \citet{Merlin2016a} performed a multi-step procedure with the program GALFIT \citep{Peng2002} in all the images, to subtract the light contribution from the foreground objects and the intracluster light.

\subsection{VLT/MUSE data}
\begin{figure}
	\includegraphics[width=\columnwidth]{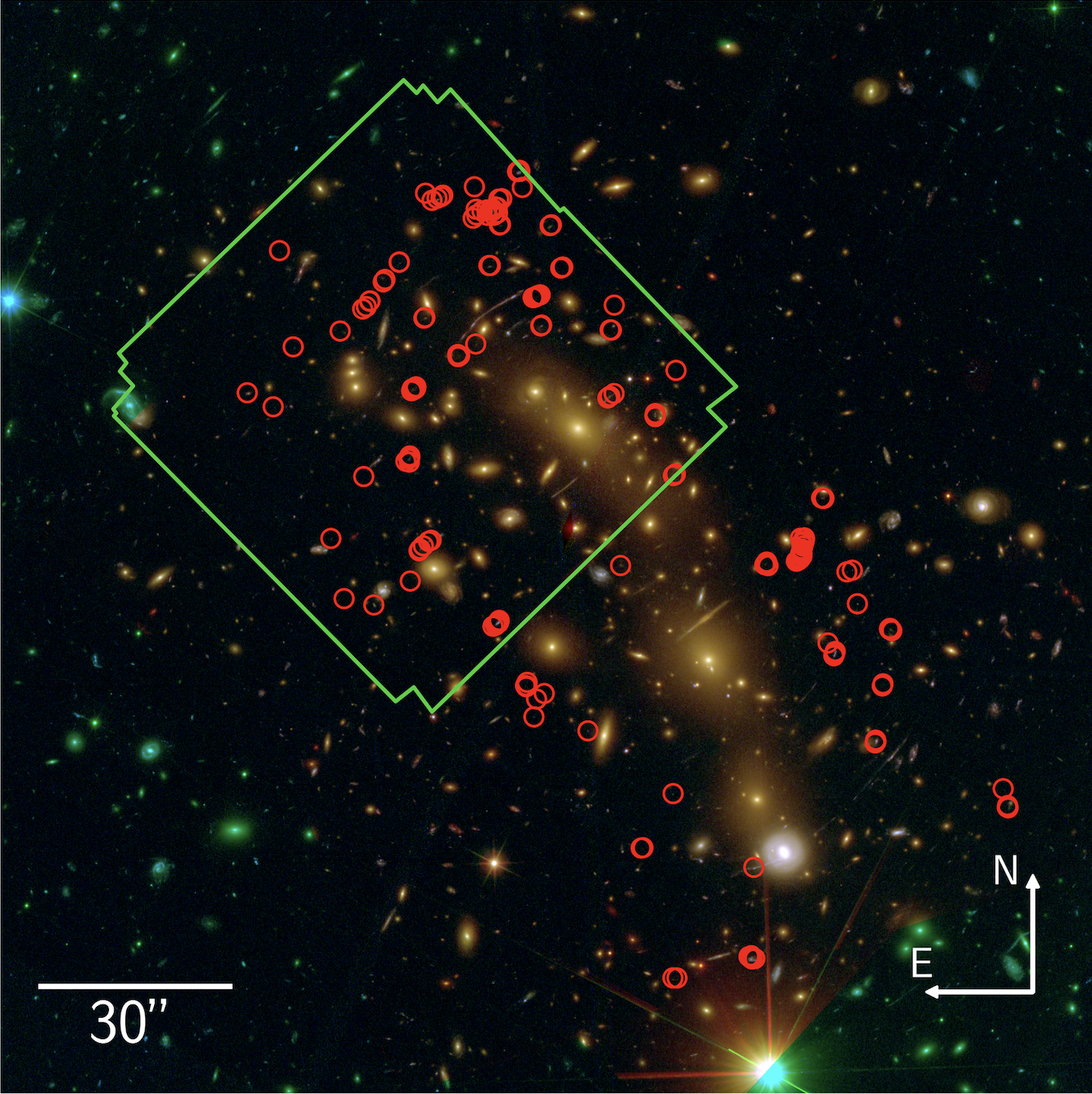}
    \caption{RGB image (R: F105W+F110W+F125W+F140W+F160W, G: F606W +F625W+F775W+F814W+F850LP, B: F435W+F475W) of MACS~J0416 showing in green the north-east 17.1~h MUSE pointing \citep{Vanzella2021b}. Red circles mark the position of the 166 individual clumps included in our sample. They all benefit from a spectroscopic redshift measurement, and 100 of them are covered by the deep MUSE pointing.}
    \label{fig:0416RGB}
\end{figure}

Thanks to its unique properties as a gravitational lens, MACS~J0416 benefits from an excellent spectroscopic coverage, which is essential to build a robust strong lensing model. We make use of the latest ground-based Integral Field spectroscopic data, observed with MUSE at the Very Large Telescope (VLT) between November 2017 and August 2019 (Prog.ID 0100.A-0763(A), PI: E. Vanzella). These observations consist of 22.1 hours (including overheads) pointing in the north-east region of the galaxy cluster (see Fig.~\ref{fig:0416RGB}). This dataset was implemented with GTO observations taken in November 2014 (Prog.ID 094.A-0115B, PI: J. Richard), reaching a total on-sky integration time of 17.1 hours, and a final image quality of 0.6$\arcsec$. MUSE data-cubes cover a field-of-view of 1 squared arcmin, spatially sampled with $0.2'' \times 0.2''$ pixels. The wavelength range extends from $4700 \, \si{\angstrom}$ to $9350 \, \si{\angstrom}$, with a dispersion of $1.25\,\si{\angstrom.pix^{-1}}$, and a spectral resolution of $\sim 2.6\,\si{\angstrom}$ approximately constant across the entire spectral range. With these data and extending the catalogs by \citet{Caminha2017} and \citet{Richard2021}, \cite{Vanzella2021b} identified and measured the redshift of 48 background sources, with $0.9 < z < 6.2$, lensed into 182 multiple images, all of them spectroscopically confirmed. 

\subsection{Lensing model}
The strong lensing model used in this work is described in \cite{Bergamini2021} and it is based on the spectroscopic catalog developed by \cite{Vanzella2021b} and described above. The total mass distribution of MACS~J0416 was accurately reconstructed, with a root-mean-square displacement of only 0.40$\arcsec$ between the observed and model-predicted positions of the 182 multiple images. This result has been possible thanks also to the identification of 171 cluster galaxy members (80\% of which spectrosopically confirmed) and to the measurement of the internal velocity dispersion for 64 of them, allowing to estimate the contribution of their subhalo mass components via the Faber-Jackson scaling relation \citep{Bergamini2019,Bergamini2021, Bergamini2023}. They also quantify the uncertainties, in different locations, on the magnification maps that can be derived from the strong lensing model. This is a fundamental step in the study of lensed high-$z$ sources, due to the fact that the local magnification factor is essential to infer the intrinsic properties from the observed ones. Since our sources are compact, when we mention the magnification factor of a clump, we refer to the average of the magnification maps from \citet{Bergamini2021}, in a region of $2 \times 2$ pixels and centered on the clump. 

\subsection{Clumps identification}
The procedure of clumps identification is detailed in \citet{Mestric2022} (hereafter, \citetalias{Mestric2022}), who made use of the same \textit{HST} PSF matched images exploited in this work. Summarizing, the clumps have been identified, and their emission deblended from that of the underlying diffuse host galaxy, by smoothing a cutout image centered on each clump through a boxcar filter with the {\tt IRAF} \citep{Tody1986} task {\tt MEDIAN}. The size of the smoothing box represents a key factor in this process, and \citetalias{Mestric2022} optimized it by assuming a size $\sim \! 2$-$3$ times larger than each source, resulting typically on 21-31 pixels. Then, the smoothed image is subtracted from the original cutout to obtain a high-contrast image. With automatic \citep[{\tt SExtractor v2.24},][]{Bertin1996} and visual inspection, it is possible to efficiently identify the individual clumps. This technique has been used in the literature in different fields, from the study of non-lensed clumps \citep{Guo2015} to the subtraction of the contaminating host galaxy before modelling ultra-compact dwarfs \citep{Norris2011}. \citetalias{Mestric2022} found that while the majority ($\sim 70\%$) of the lensed galaxies host 1-2 clumps, there are systems harbouring up to 15-16 clumps. For more details on the distributions of clumps in number per system and redshift, we refer the readers to the \citetalias{Mestric2022} paper.

\subsection{Sample} 
The resulting \citetalias{Mestric2022} sample we analyze is composed of 166 star-forming clumps belonging to 67 galaxies strongly lensed by MACS~J0416. The magnification factors of the clumps in our sample are distributed from $\simeq \! 2$ to $\simeq \! 82$, with a median value of approximately 4.6. About 80\% of the clumps are magnified by a factor $<10$. Approximately 70\% (48 over 67) of the galaxies has multiple images: for them, we consider the brightest system, that is the one with the largest value of the magnification factor, from the lensing model developed by \cite{Bergamini2021} and described above. When possible, we check the consistency of the presented results with those relative to the not considered multiple images.
All the clumps in the sample benefit from a spectroscopic redshift measurement, and they cover a redshift range from $z \sim 2$ to $\sim 6.2$, with peaks around redshift 2, 3.5 and 6, due to clumps hosted by the same galaxy, or by different galaxies at approximately the same redshift, probably belonging to the same group or structure. Among the 166 clumps in the sample, 100 are included in the deep MUSE pointing of \cite{Vanzella2021b}, shown in Fig.~\ref{fig:0416RGB}.

\section{Photometric $\beta$ slopes}
In this section we describe how we measure the fluxes relative to each clump, and the methodology exploited to measure the respective photometric $\beta$ slopes. We then discuss how we estimate the robustness of the resulting measurements. After that, we present our $\beta$ slopes measurements and their uncertainties. At the end of the section, we compare our results with those obtained by adopting the fluxes measured by the {\tt ASTRODEEP} collaboration \citep{Merlin2016a, Castellano2016} for a subsample of 48 in-common objects.

\subsection{Photometric measurements}
We measure the photometric properties of clumps on the images reduced by the {\tt ASTRODEEP} collaboration, where the intracluster light (ICL) and the foreground cluster members are subtracted and the images in each filter are PSF matched to the resolution of the \textit{H160} one. 
We make use of the A-PHOT software \citep{Merlin2019}, developed to perform aperture photometry on astronomical images, which allows one to obtain multiple measurements within different circular or elliptical apertures, and to estimate and subtract the local background sky. A-PHOT computes the total flux within an aperture by summing up the flux of the pixels entirely included in the aperture, and dividing in $n \times n$ sub-pixels those crossed by the border, with $n$ fixed by the user, and iterating the procedure. The local background is estimated through a recursive algorithm with a clipping procedure, considering the mean value of the pixels within an annulus centered on the aperture. 

We measure the fluxes relative to each clump by considering circular apertures of diameter $0.27\arcsec$, centered on the center of each clump (see Table~1 of \citetalias{Mestric2022} for the coordinates), and with the local background subtraction option implemented. The output magnitudes are computed as $m_i = -2.5 \log f_i + zp_i$, where $i$ denotes each filter, $f_i$ are the fluxes measured with A-PHOT and $zp_i$ are the relative zeropoints.

\subsection{Photometric $\beta$ slopes}
We measure the UV-continuum $\beta$ slopes through the relation \citep{Castellano2012} 
\begin{equation}
\label{eq:fitCastellano}
    m_i = -2.5(\beta + 2) \log(\lambda_i) + c \, \, \, ,
\end{equation} 
where $m_i$ is the measured magnitude in the $i$-th filter and $\lambda_i$ is the corresponding wavelength, assumed to be the pivot wavelength\footnote{defined, for each filter, as $\lambda_p = \sqrt{\frac{I(\lambda) \diff \lambda}{I(\lambda) \lambda^{-2} \diff \lambda}}$, where $I(\lambda)$ is the response of the filter.} of each filter reported to rest-frame, i.e., divided by a factor $(1+z)$. We fit the data with a weighted least squares technique, where the weights, $w_i$, depend on the magnitudes uncertainties, $\epsilon_{m, \, i}$ as $w_i = \epsilon_{m, \, i}^{-2}$. We correct the seven bands for Milky Way reddening by adopting the \citet{Cardelli1989} reddening law with $R_V=3.1$ \citep{Schlafly2011, ODonnell1994}.

\begin{figure*}
	\includegraphics[width=\textwidth]{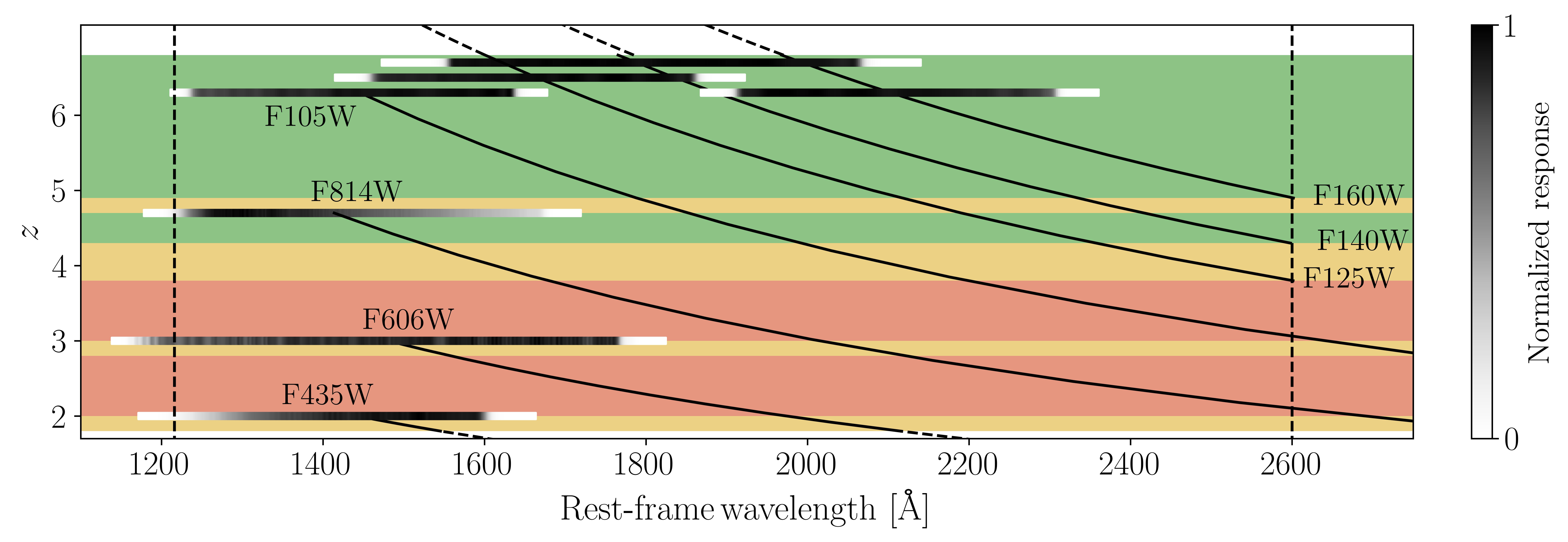}
    \caption{ Scheme of the rest-frame UV coverage of each available HST filter, as a function of the redshift (vertical axis). The UV limits \citep[$1216-2600$ \AA,~][]{Calzetti1994} are displayed with vertical dashed lines.
    Each filter is entirely represented at the respective $z=z_\mathrm{sup}$ with the intensity of the color proportional to the normalized response (colorbar on the right), while a solid line shows the position of its pivot wavelength as a function of the redshift. In this representation, a filter is defined as rest-frame UV in the redshift interval where the solid line is drawn. That is, fixing a value of the redshift (i.e., tracing an horizontal line), the included filters are those whose solid lines are crossed. The coloured backgrounds indicate the corresponding number of included filters: two (red), three (yellow) or four (green). For the F125W, F140W, and F160W filters, the dashed end of the lines indicates that the filter could potentially be further exploited to higher redshifts, out of the range of our sample. The solid lines relative to the F814W and F105W filters extend over the $2600 \, \si{\angstrom}$ limit because we relaxed the selection criteria, as justified in the main text.
    The filters and the number of clumps of the sample included in each redshift interval are also summarized in Table~\ref{tab:bands}.}
    \label{fig:bands2}
\end{figure*}

For a given value of the redshift, only the fluxes measured in filters that are rest-frame included in the UV interval can be exploited to fit Eq.~\ref{eq:fitCastellano}. We adopt the following criteria to select the redshift range in which each filter can be exploited: we measure the redshift limits, $z_\mathrm{inf}$ and $z_\mathrm{sup}$, for each filter by considering the redshift values such that the pivot and the minimum wavelengths, $\lambda_{p}$ and $\lambda_{min}$, are included in the $1250$-$2600\, \si{\angstrom}$ range \citep{Calzetti1994}, as 
\begin{equation}
\label{eq:condition1}
    z_\mathrm{inf} = \dfrac{\lambda_{p}}{2600 \, \si{\angstrom}} - 1
\end{equation}
\begin{equation}
\label{eq:condition2}
    z_\mathrm{sup} = \dfrac{\lambda_{min}}{1216 \, \si{\angstrom}} - 1 \, .
\end{equation}
In particular, we use $\lambda_{min}$ towards the $1250$ \AA\, limit to avoid the possible inclusion in the filter of the Ly$\alpha$ emission line, that can significantly affect the measured magnitude.
Considering that the clumps range from redshift 2 to 6, we divide our sample into seven redshift intervals, as described in Table~\ref{tab:bands} and showed in Fig.~\ref{fig:bands2}. Each interval differs for the number and the kind of exploitable filters.
By following the described criteria, we find that in the redshift interval between 2.8 and 3.0, only one filter (\textit{I814}) can be exploited, and thus no UV $\beta$ slope measurement is possible for 11 clumps. Thus, we relax the conditions and include the \textit{V606} band, whose transmission at $1216 \, \si{\angstrom}$ is $<10\%$, and the \textit{Y105} band. We test the results by measuring the slopes with only the \textit{V606}-\textit{I814} and \textit{I814}-\textit{Y105} pairs in the same redshift interval, obtaining fully consistent results, differing only by a few percent. Thus, we present in the following the three-magnitude slope for the 11 clumps included in the redshift interval between 2.8 and 3.0. Similarly, we extend the redshift range of the \textit{I814} filter to lower values, to cover also 9 clumps at $z \sim 1.9$. We tested the consistency of the results with the only-two filter fit. In this case, the choice is supported and justified by the asymmetric shape of the \textit{I814} response, peaking at $\lambda_\mathrm{peak} < \lambda_{p}$.


\begin{table}
\centering
\begin{tabular}{@{}rccc@{}}
\toprule
\toprule
Redshift interval & \# filters & Filters & \# clumps \\
\midrule
$z < 2.0$ & 3 & \textit{B435, V606, I814} & 9 \\
$2.0 < z < 2.8$ & 2 & \textit{V606, I814} & 40 \\
$2.8 < z < 3.0$ & 3 & \textit{V606, I814, Y105} & 11 \\
$3.0 < z < 3.8$ & 2 & \textit{I814, Y105} & 51 \\
$3.8 < z < 4.3$ & 3 & \textit{I814, Y105, J125} & 25 \\
$4.3 < z < 4.7$ & 4 & \textit{I814, Y105, J125, JH140} & 4 \\
$4.7 < z < 4.9$ & 3 & \textit{Y105, J125, JH140} & 2 \\
\multicolumn{1}{l}{$\, 4.9 < z $} & 4 & \textit{Y105, J125, JH140, H160} & 24 \\
\bottomrule
\end{tabular}
\caption{Summary of the HST filters used to fit Eq.~\ref{eq:fitCastellano} in each redshift interval, following the criteria described in Eqs.~\ref{eq:condition1} and \ref{eq:condition2}. For each redshift interval, we report the number of included rest-frame UV filters, their names, and the respective number of clumps in our sample.}
\label{tab:bands}
\end{table}

Hence, the $\beta$ slope for each clump is computed by fitting Eq.~\ref{eq:fitCastellano} with two, three, or four magnitudes relative to the rest-frame UV filters. The procedure and the parameters used to measure the magnitudes can play a major role in the resulting $\beta$ slopes. In the following, we describe the software we use and the procedure we follow.  

\subsection{Robustness of the photometric and $\beta$ slopes measurements}
We choose this combination of parameters (circular apertures of diameter $0.27\arcsec$, centered on the center of each clump, with the local background subtraction option implemented) after performing several tests on the real clumps and on 50 mock clumps we injected in the images as PSF functions, in locations similar to those of the real clumps. In detail, we distribute them in the outer and inner regions of the cluster, to see the possible residuals from the intracluster light removal process, in isolated positions and angularly close to a bright object, to quantify the contribution of the contamination of foreground galaxies. For instance, in the latter case, we put the mock clump at the same angular distance to the contaminant as that of the real clump, but in an opposite direction, to avoid the real clumps to contaminate the simulation. The resulting locations of the 50 mock clumps are shown in the Appendix \ref{app:extra_sim}, in Fig.~\ref{fig:sim_loc}.

We test different apertures, from $0.2 \arcsec$ to $0.54\arcsec$ in diameter, to switch on and off the A-PHOT local background estimation, and to manually fit and subtract with GALFIT the surface brightness distribution of a foreground contaminant and the background level. For each mock clump, we associate a ``true'' ($\beta_{true}$) UV slope value extracted from an uniform distribution in the $\left[ -3, 0 \right]$ interval, and a $m_{I814}$ magnitude in the \textit{I814} band sampled from a Gaussian distribution with $\mathrm{mean} = 28.1$ and $\sigma=1.1$, resulted by fitting the magnitude distribution of the clumps in our original sample. From $\beta_{true}$ and $m_{I814}$ it is possible to uniquely assign $m_{Y105}$, the corresponding magnitude in the \textit{Y105} band. Analogously, in the next steps we assign the respective magnitudes in the \textit{J125} and in the \textit{JH140} filters. At the end of each step, we measure the magnitude of the mock clumps with A-PHOT, exploiting in sequence the two, three and four available bands.
 
\begin{figure*}
	\includegraphics[width=\textwidth]{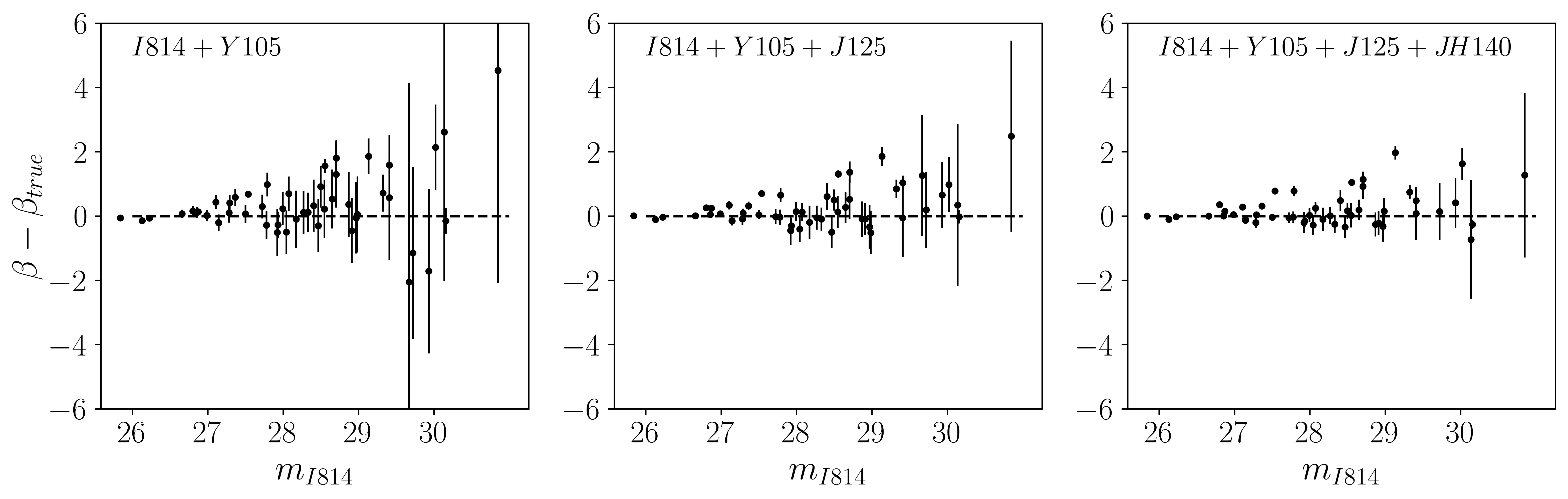}
    \caption{Difference between the measured and the injected $\beta$ slopes ($\beta-\beta_{true}$) as a function of the magnitude $m_{I814}$, in the case with aperture of diameter 0.27\arcsec, when $\beta$ is measured using the magnitudes in two (left), three (center) or four (right panel) filters. The scatter around zero increases with the faintness of the sources, because of the less precise photometric measurements, but also with decreasing number of exploited bands, being however mostly consistent with zero within a $1\sigma$ uncertainty.}
    \label{fig:sims_results}
\end{figure*}

We check how the photometric measurements change when modifying the aperture, the background subtraction, and the contaminants subtraction. We find good agreement between the measured $\beta$ slopes with different apertures and the input $\beta_{true}$ values. We show, in Fig.~\ref{fig:sims_results}, the residuals ($\beta-\beta_{true}$) as a function of $m_{I814}$, in the selected case with aperture 0.27\arcsec-diameter. The scatter along the zero-residual line increases with the faintness of the sources, due to the less precise photometric measurements, but also with decreasing number of exploited bands, although remaining consistent with zero within $1\sigma$ uncertainties: we describe extensively how we estimated the uncertainties on the measured $\beta$ values in Subsect.~\ref{subs:unc_beta}. 
We repeat the experiment by extracting the photometry from a circle with a larger aperture of diameter 0.4\arcsec and find that the measured $\beta$ slopes are systematically redder (i.e., larger values) than those obtained with aperture of diameter 0.27\arcsec, of typically $\Delta \beta \sim 0.2$, corresponding to $\sim \! 10\%$. There are two main reasons that can explain this result. Increasing the aperture means that a larger fraction of the light from the diffuse host galaxy is included, and it has typically a redder slope. Moreover, it also includes a larger light contribution from the ICL (or from the residuals of its subtraction) and from foreground contaminants. It is the case of several clumps in our sample, which are located in positions angularly close to a red galaxy. For this subsample, we model and then subtract the surface brightness distribution of the contaminant with GALFIT in the different bands involved in the $\beta$ slope measurement. Then, we repeat the measurement on the cleaned images. With this procedure (0.4\arcsec-diameter aperture, A-PHOT sky subtraction off, contaminants subtracted with GALFIT) we recover $\beta$ values consistent with those obtained in our best case, which is 0.27\arcsec-diameter aperture and A-PHOT sky subtraction on. Finally, we adopt and present in the following the results obtained with this last configuration, to maintain the same aperture and being consistent with \citetalias{Mestric2022}.

\subsection{$\beta$ slopes measurements}
\begin{figure*}
	\includegraphics[width=\textwidth]{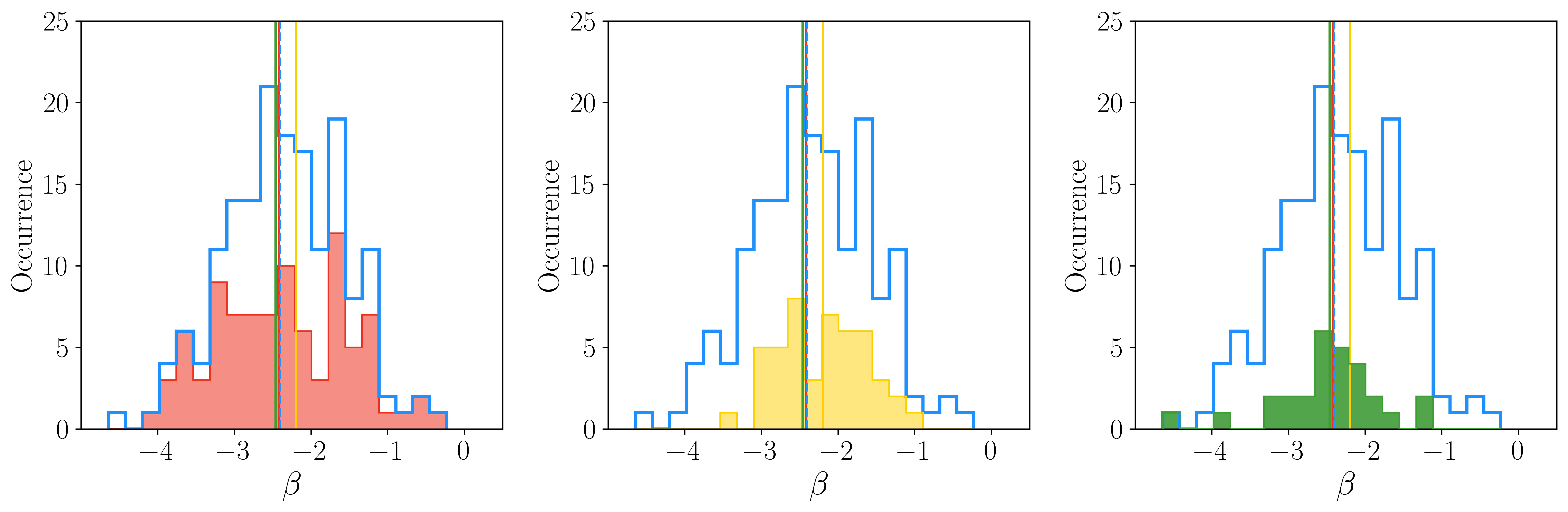}
    \caption{Distribution of the photometric $\beta$ slopes measured (blue empty histogram), and split into subsamples with two (red), three (yellow), or four (green) available photometric measurements. The vertical lines represent the median value relative to each color-coded subsample, while the blue dashed one represents the median value of the entire sample. }
    \label{fig:beta_hist}
\end{figure*}

The resulting UV-continuum $\beta$ slopes are shown in Fig.~\ref{fig:beta_hist}. The $\beta$ slopes distribution of our sample of individual clumps shows a median value of $\sim -2.4$, with a standard deviation of $0.78$. The low median value is expected, since UV bright clumps in high-$z$ galaxies are well-known sites of star formation \citep[e.g.,][]{Bournaud2014, Zanella2015, Zanella2019, Mestric2022}. Thus, they are populated by young, massive OB stars, whose spectrum strongly emits in the UV, resulting in a blue $\beta$ slope. 
There are several objects populating the tails of this distribution: 4 objects have a very red ($\beta>-1$) and 50 a very blue ($\beta < -2.7$) slope. We will discuss these extreme clumps in Subsect.~\ref{subs:blue}. We observe that the majority of extremely blue slopes was measured with 2-magnitudes fits, that are more affected by systematics, but also 8 clumps with 3-magnitudes and 8 with 4-magnitudes fits are included. The median $\beta$ slopes of each subsample are $-2.45$, $-2.20$, and $-2.40$, when two, three, and four magnitude measurements are exploited in the fit, respectively.

\subsection{Uncertainty estimates on $\beta$}
\label{subs:unc_beta}
The uncertainties associated to the $\beta$ slopes primarily depend on the magnitude uncertainties (Fig.~\ref{fig:sims_results}). A-PHOT computes the uncertainty associated to the flux of the object \textit{obj} through the so-called ``CCD Equation" \citep{Mortara1981}, which assumes the form
\begin{equation}
   \sigma_{obj} = \sqrt{ \sum_{i=1}^{N_{pixels}} \mathrm{rms}^2_{i, \, obj} + \frac{f_{obj}}{G}} 
\end{equation}
when the root-mean-square (rms) map is considered. Here, $f_{obj}$ is flux received associated to the object and $G$ is the gain.

We adopt these estimates as the $1\sigma$ uncertainty on the flux measurements, and use them to compute $\epsilon_m$, the uncertainty relative to the corresponding magnitude value. Then, we use a bootstrap technique to estimate the uncertainties on the $\beta$ slopes, by fitting, via Eqn.~\ref{eq:fitCastellano}, $10^4$ sets of magnitudes randomly extracted from a Gaussian distribution centered on the measured values in each filter and with $\sigma=\epsilon_m$. We assess the 16$^\mathrm{th}$ and the 84$^\mathrm{th}$ percentile values of the resulting $\beta$ distribution as the lower and upper $1\sigma$ uncertainties, respectively. 
We notice that clumps with 2-magnitudes fits generally lie at the lowest redshifts (Table~\ref{tab:bands}) and consequently they dominate the bright end of the magnitude distribution of clumps. It results in $\beta$ slopes of bright objects (i.e., with accurate flux measurements) with only two photometric measurements to have smaller uncertainties than those with three or four photometric points, which do not reflect the systematics that affect such two-point $\beta$ slope measurements. To take into account systematics, we apply to all the 2-magnitudes fits the same bootstrap technique, but extracting a random value from a Gaussian distribution centered on the measured values in each filter and with $\sigma = 3\epsilon_m$. The uncertainties on the 2-point $\beta$ slopes estimated in this way are consistent with those estimated for the 3- and 4-point fits.

\subsection{Comparison with ASTRODEEP}
The objects in the MACS~J0416 field have been extensively detected and characterized by the {\tt ASTRODEEP} collaboration \citep{Merlin2016a, Castellano2016}, with a $5\sigma$ depth in the range 28.5-29.0 AB in 2 PSF-FWHM ($=$~0.2\arcsec) aperture. \citetalias{Mestric2022} cross-matched our sample with the {\tt ASTRODEEP} catalog, finding 48 in-common objects. For this subsample, we consider the fluxes measured by the {\tt ASTRODEEP} collaboration in each HFF filter\footnote{\href{http://astrodeep.u-strasbg.fr/ff/}{http://astrodeep.u-strasbg.fr/ff/}}, and apply our pipeline to measure the associated $\beta$ slope, using the same approach and band selection described above. The comparison with our $\beta$ slopes is shown in Fig.~\ref{fig:comparison_AD}.

We find in general good agreement between the resulting $\beta$ slopes. The {\tt ASTRODEEP} slopes are systematically redder (median $\Delta \beta \sim 0.24$), but consistent with the 1:1 relation given the average uncertainty of approximately 0.27. Moreover, the redder {\tt ASTRODEEP} slopes can be explained by the fact that the fluxes are measured in larger apertures and that the catalog is mainly composed of galaxies, and not of individual clumps. In almost all the cases, each {\tt ASTRODEEP} object is composed by multiple clumps of our sample plus their diffuse host. For these cases, \citetalias{Mestric2022} associated the {\tt ASTRODEEP} object to the brightest clump of the group, but the correspondence is not exactly one-to-one, and also the centroids may be slightly shifted. The agreement of the results, within less than $1\sigma$ on average, represents an important consistency test for the robustness of our flux measurements, given that the {\tt ASTRODEEP} fluxes are calculated with a different approach but exploiting exactly the same images, with subtracted ICL and PSF-matched to the \textit{H160} band.

\section{Spectroscopic $\beta$ slopes}
\label{sect:spectroscopic}
Similarly to \citet{Calzetti1994}, we exploit ten spectral windows in the rest-UV range, from 1200 to 2600 \AA\, (see Table~\ref{tab:windows}), to measure the spectroscopic $\beta$ slope of the 100 clumps of our sample included in the MUSE pointing. These intervals are properly designed to remove from the fitting the main absorption and emission lines, as well as the strong telluric absorption residuals, that could bias the measurement of the continuum slope. For each window, we measure the integrated flux and associated uncertainty, correct the them for Milky Way reddening (\citealt{Cardelli1989} reddening law with $R_V=3.1$, \citealt{Schlafly2011, ODonnell1994}), and then fit Eqn.~\ref{eq:fitCastellano}. Depending on the redshift of each clump, the MUSE wavelength coverage reported to rest-frame allows us to employ a different number of windows, that is, a different number of flux values for the fit. The majority of clumps can be fit with $\gtrsim 6$ windows, and we exclude the clumps whose spectrum covers less than three windows, as it happens for $z>5.7$. Hence, we extract the spectrum for 87 clumps in our sample. 

\begin{table}
\centering
\begin{tabular}{@{}ccc@{}}
\toprule
\toprule
Window number & Wavelength range [\AA] & Redshift range \\
\midrule
1 & 1268 - 1284 & $z \gtrsim 2.6$ \\
2 & 1309 - 1316 & $2.5 \lesssim z \lesssim 6.1$\\
3 & 1360 - 1371 & $2.4 \lesssim z \lesssim 5.8$\\
4 & 1407 - 1515 & $2.1 \lesssim z \lesssim 5.6$\\
5 & 1562 - 1583 & $1.9 \lesssim z \lesssim 4.9$\\
6 & 1677 - 1725 & $z \lesssim 4.4$ \\
7 & 1760 - 1833 & $z \lesssim 4.3$ \\
8 & 1866 - 1890 & $z \lesssim 4.0$ \\
9 & 1930 - 1950 & $z \lesssim 3.8$ \\
10& 2400 - 2580 & $z \lesssim 2.9$\\
\bottomrule
\end{tabular}
\caption{Rest-frame UV ten spectral windows used to measure the spectroscopic $\beta$ slopes and the relative redshift range in which they can be exploited.}
\label{tab:windows}
\end{table}

We extract the spectra of each source by fixing circular apertures of 0.4\arcsec diameter, centered on each clump. Since most of the sources in our catalog, in particular at higher redshifts, are very faint, we estimate the spectra signal-to-noise ratio 
($S/N$), and we keep only those with $S/N \gtrsim 2$. After this selection, we measure the spectroscopic $\beta$ slope for 37 clumps. They are distributed between redshift 1.99 and 3.29, corresponding to 6-9 exploited spectral windows. 

\subsection{Comparison with photometric $\beta$ slopes}
The spectroscopic $\beta$ slopes are on average redder than the photometric ones, with a median $\Delta \beta = \beta_\mathrm{spec}-\beta_\mathrm{phot} \sim 0.7$.
There are different factors that contribute to it. The main one is the contamination from some red foreground objects, the BCG, and the ICL, which are not subtracted in the MUSE datacube, unlike the HST images. Additionally, we estimated the photometric slopes using magnitudes extracted from 0.27\arcsec-diameter apertures, increased, given that the MUSE observations are seeing limited, to 0.4\arcsec-diameter for the spectroscopic ones. The larger aperture enhances the effect of contaminants, as we observe in photometric measurements, where the median difference between slopes measured with 0.4\arcsec and 0.27\arcsec-diameter aperture is 0.21. This effect can be seen in Fig.~\ref{fig:specVSphot_position}, where we show the difference between the photometric and spectroscopic $\beta$ slopes as a function of the position in the sky (i.e., of the presence of close by contaminants) and of the redshift. Thus, we discard the clumps with angular distance smaller than $5$\arcsec\, from the closest foreground red galaxy, reducing the spectroscopic sample to 27 clumps with reliable both photometric and spectroscopic $\beta$ slopes, and the difference between them is reduced to $\Delta \beta \sim 0.3$.

\section{Results and discussion}
\subsection{Comparison with galaxy-integrated $\beta$ slopes}
\begin{figure}
	\includegraphics[width=\columnwidth]{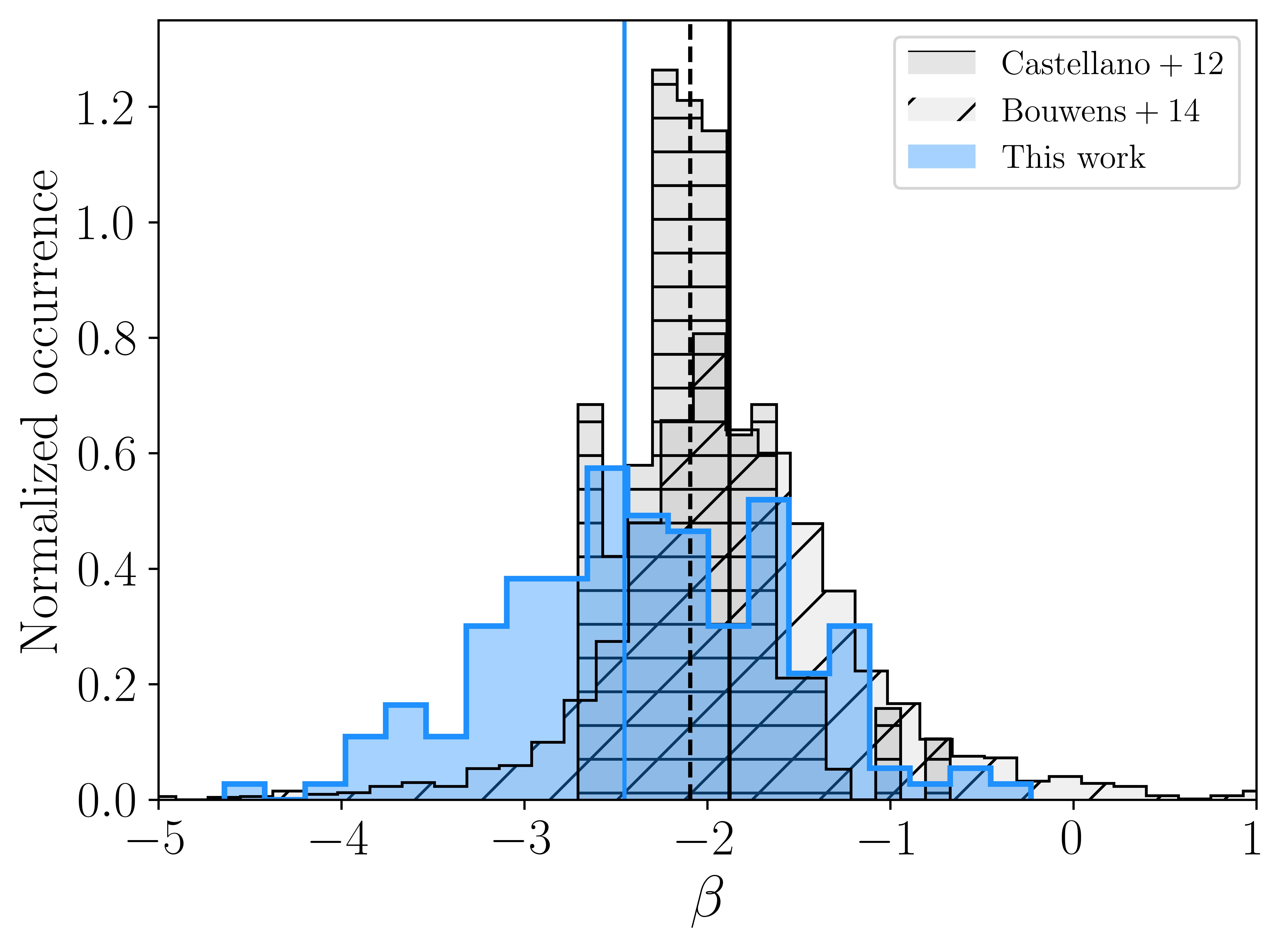}
    \caption{Distribution of the measured photometric $\beta$ slopes. Our sample of clumps (in blue) is compared with two samples of high-$z$ galaxies, from \citet{Castellano2012} and \citet{Bouwens2014} (hatched black histograms). The vertical lines represent the median values of the distributions, respectively in blue, dashed black and solid black.}
    \label{fig:beta_hist_comp}
\end{figure}

We compare the resulting photometric $\beta$ slopes for our sample of individual clumps with those of galaxies at $z\sim 4$, color-selected from the GOODS-ERS WFC3/IR dataset \citep{Giavalisco2004} and HUDF WFC3/IR dataset \citep[e.g.][]{Oesch2010} by \citet{Castellano2012}, and with a sample of galaxies from $z=4$ to 8 from \cite{Bouwens2014}. The comparison is shown in Fig.~\ref{fig:beta_hist_comp}. The samples of galaxies reveal that most of them have blue UV slopes, with distributions peaking around $\beta \sim -2$, with some red slope interlopers ($\beta \gtrsim -0.5$). This has been interpreted as a suggestion of low dust environment in high-$z$ galaxies \citep[e.g.,][]{Dunlop2013}. The $\beta$ slopes distribution of our sample of individual clumps shows a bluer median value of $\sim -2.4$, that is consistent with different scenarios. Firstly, it confirms that clumps are sites of intense star formation \citep[e.g.,][]{Bournaud2014, Zanella2015, Zanella2019, Mestric2022}, and are populated by young, massive OB stars, whose spectrum strongly emits in the UV. But, it can also point out different features between the host galaxies and their clumps. The same result can be indeed reproduced by assuming that the dust at the location of the clumps is lower than the average extinction of the galaxy. This implies that assuming a similar extinction for the clumps and the host would result in an overestimate of the clumps SFR. Also, a lower metallicity at the clumps location or a different IMF or SFH can explain the bluer median $\beta$ slope, and would have a crucial role in estimating the age of the clumps.

\subsection{Trends with $M_{UV}$ and redshift}
\begin{figure*}
	\includegraphics[width=\textwidth]{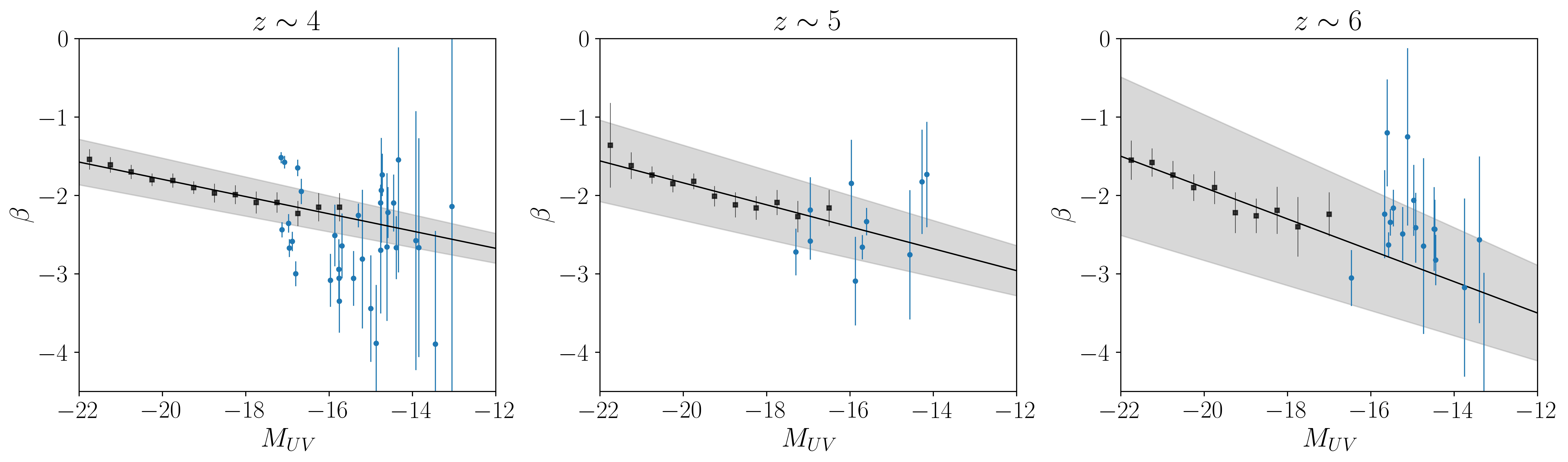}
    \caption{Measured $\beta$ values as a function of the UV magnitude $M_{UV}$, in the $z \sim 4$ (left), $z \sim 5$ (center), $z \sim 6$ (right) redshift bin. The blue points, with $1\sigma$ uncertainties, are the clumps of our sample, while the bi-weighted binned mean of the sample of integrated galaxies from \citet{Bouwens2014} are shown in black. The black line represents the best-fit relation measured from this latter sample. The relation still holds at much fainter magnitudes, suggesting that star-forming clumps and their hosts follow the same relation.}
    \label{fig:beta-Muv}
\end{figure*}

It has been shown in previous studies that the mean UV-continuum slope of galaxies shows a dependence on the UV luminosity, with an almost identical slope with redshift ranging from 4 to 10 \citep{Bouwens2014, Yamanaka2019, Cullen2022}. It suggests that UV-faint galaxies are typically younger, less metal-enriched, and less dust-obscured than their brighter analogs \citep[e.g.,][]{Rogers2013, Bhatawdekar2021}. \\
We show the results for our sample in Fig.~\ref{fig:beta-Muv}, with superimposed the relation found by \citet{Bouwens2014}. The UV magnitudes for our sample, $M_{UV}$, have been measured as the geometric mean absolute magnitude of each clump in all the \textit{HST} bands that contribute to its UV-continuum $\beta$ slope determination. The adoption of the geometric mean prevents one to give too much weight to the bluer or redder bands in measuring $M_{UV}$, that could artificially introduce a $\beta$-$M_{UV}$ correlation \citep{Bouwens2012}. The observed magnitudes are converted to intrinsic through the local magnification factor, and then to absolute magnitudes via the luminosity distance, measured with the assumed cosmological model. We show the results dividing our sample in redshift bins, as $z \sim 4$, $z \sim 5$, and $z \sim 6$. 

The clumps in our sample are consistent with the \citet{Bouwens2014} reference relation within $2.3\sigma$ ($z \sim 4$), $1.2\sigma$ ($z \sim 5$) and $0.5\sigma$ ($z \sim 6$)\footnote{We measured the consistency values, $\eta$, as $$\eta = \frac{|\beta-\beta_\mathrm{B14}|}{\sqrt{\epsilon_\beta^2 + \epsilon_{\beta_\mathrm{B14}}^2}} \, \, ,$$ where $\beta$ and $\epsilon_\beta$ are the measured slopes and their uncertainties, while the subscript B14 denotes the same quantities derived from the \cite{Bouwens2014} best-fit relation.}, and with a median scatter of $\Delta \beta \sim $ 0.26, 0.15, and 0.35, respectively. The increasing consistency at higher redshift is mainly due to the larger uncertainty of the $z\sim 6$ relation and to the increasing median uncertainties on the $\beta$ measurements, but it may also suggest different evolutionary schemes. 
In fact, it may convey that clumps are bluer than their hosts especially at lower redshifts. It suggests that, at high-$z$, clumps and host have more similar stellar populations, dust content, SFH, and then the host changes properties over time more significantly than clumps do, presenting, at lower redshift, a dust rich environment and an evolved stellar population (resulting in a redder $\beta$ slope) while the clumps maintain their blue $\beta$ slopes thanks to the continuous star formation activity.
Alternatively, due to the lack of resolution and to their faintness, high-$z$ galaxy might rarely be resolved, and thus at the highest redshifts galaxies and isolated clumps can be misidentified or represent the same physical objects.

The reference relation has been measured for galaxies with $M_{UV}$ ranging from $-22$ to $-16$, while our sample of clumps covers $M_{UV}$ values between $-18$ and $-12$. It suggests that star-forming clumps follow the same $\beta$-$M_{UV}$ relation of their host galaxies, and that it can be extended to fainter magnitudes (Fig.~\ref{fig:beta-Muv}). 

\begin{figure}
	\includegraphics[width=\columnwidth]{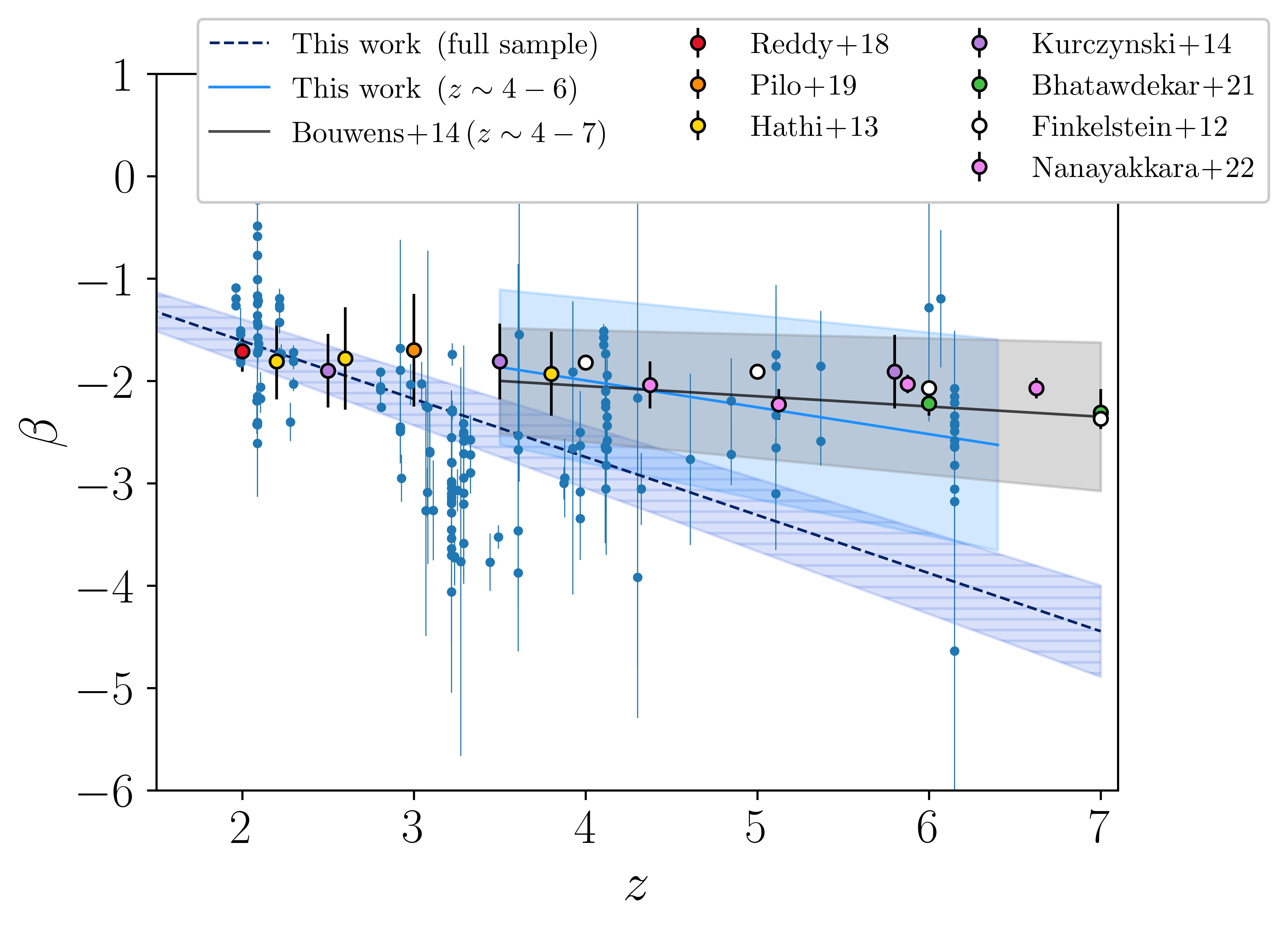}
    \caption{Measured $\beta$ values as a function of the redshift $z$, with $1\sigma$ uncertainties. We report three different fits to a possible $\beta-z$ relation: in dashed blue the best-fit weighted relation of the entire sample of clumps, in black the relation found by \citet{Bouwens2014} for a sample of integrated galaxies at redshift between 4 and 7, and in solid blue the best-fit weighted relation to our sample of individual clumps limited to the same redshift range. The median value of $\beta$ for galaxies at different redshifts reported in other studies is also shown \citep{Reddy2018, Hathi2013, Pilo2019, Kurczynski2014, Finkelstein2012, Bhatawdekar2021, Nanayakkara2022}}.
    \label{fig:beta-z}
\end{figure}

Several works \citep[e.g.,][]{Stanway2005, Wilkins2011, Finkelstein2012, Castellano2012, Bhatawdekar2021} also report an evolution of the measured $\beta$ slopes toward the blue with the increasing redshift, but it has not been confirmed uniformly with most recent JWST data \citep[e.g.,][]{Nanayakkara2022} and might be a result of observational limits. For the clumps of our sample, we observe a moderate reddening of the UV-continuum slopes with increasing cosmic time (Fig.~\ref{fig:beta-z}). By fitting a linear relation between $\beta$ and $z$, we measure $\beta = (-0.57 \pm 0.05) z + (-0.47 \pm 0.12)$.  We stress however that the slope is strongly constrained by the clumps at $z \lesssim 3$, whose photometric measurements (and, consequently, $\beta$) are less uncertain. We fit the relation excluding the $\beta$ measured with only two magnitudes, and find a fully consistent slope $\diff \beta/\diff z = -0.49 \pm 0.06$. Our relation shows a steeper slope than that measured by \cite{Bouwens2014} at $z \gtrsim 3.5$. If we limit our fit to the clumps in the same redshift range, we obtain a much flatter slope $\diff \beta/\diff z = -0.26 \pm 0.09$. 
We investigate the substantially different relations obtained in the two redshift intervals by assuming different subsets of the sample. When considering only the clumps in the most populated magnitude bins ($18 < M_\mathrm{UV} < -15$) or with secure $\beta$ measurement (uncertainty on $\beta<0.5$) we find the best-fit parameters which are completely consistent with those obtained when fitting the entire sample (difference smaller than the typical $10\%$ uncertainties).
We do not consider other functional analytical forms to fit the data, like higher degree polynomials, given that they would not be physically motivated. 

\subsection{Extremely blue $\beta$ slopes}
\label{subs:blue}
Particular interest is recently devoted to extremely blue slopes, approaching values $\lesssim -3$. Such blue slopes would imply non-standard physical properties of high-$z$ galaxies, and their analysis is crucial to characterize their stellar populations and put them in the context of galaxy formation and evolution. Several works have claimed the detection of robust $\beta \lesssim -2.7$ slopes for spectroscopically confirmed galaxies at high redshift \citep[e.g.,][]{Bouwens2010, Labbe2010, Zackrisson2013, Maseda2020, Jiang2020, Marques-Chaves2022}, while there is not yet any reference for individual star-forming clumps. 

\begin{figure}
	\includegraphics[width=\columnwidth]{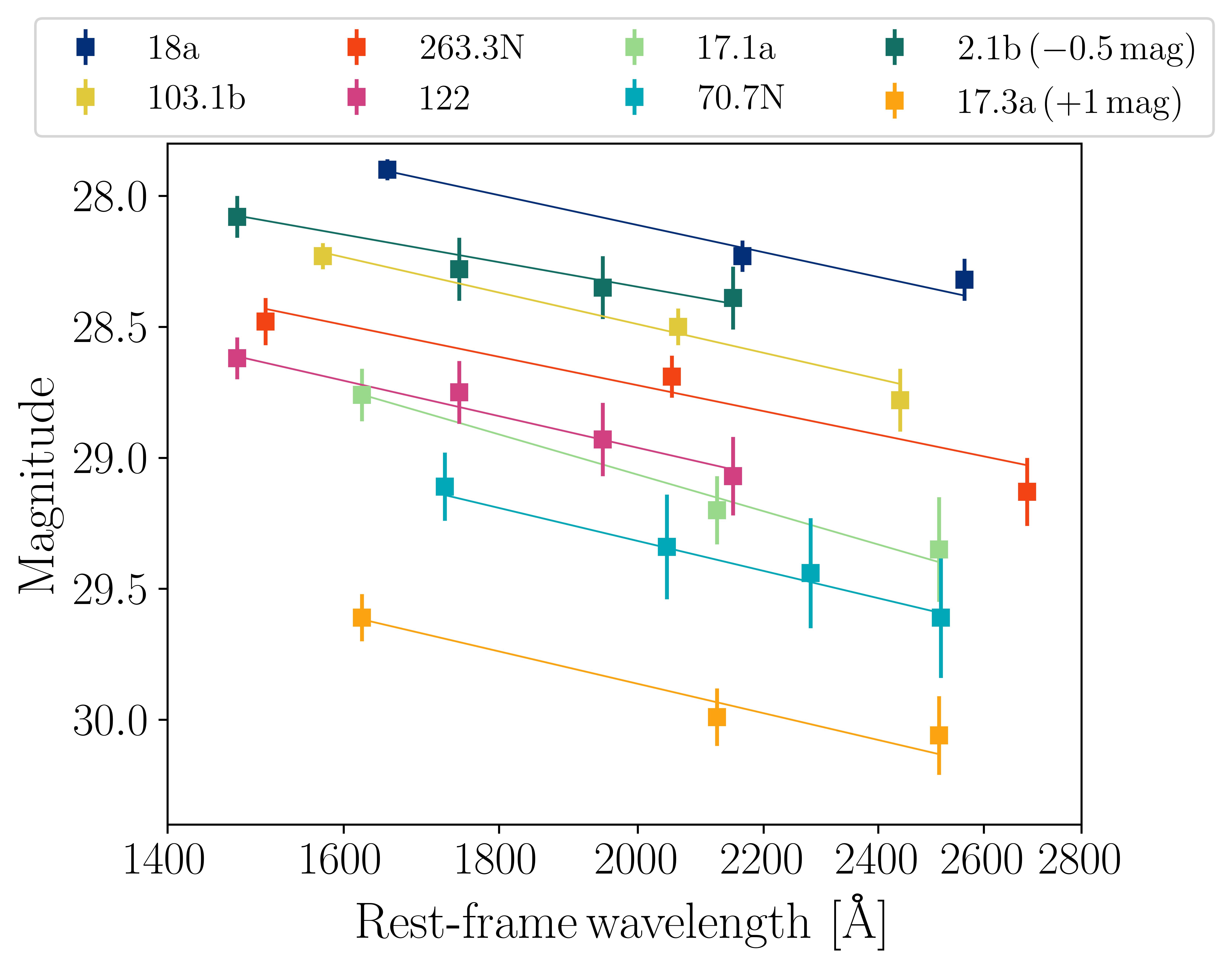}
    \caption{Measured magnitudes and the UV-continuum slope for eight selected clumps with $\beta \lesssim 2.7$, reported in Table~\ref{tab:blue_slopes}. For each clump, represented with a different color, the symbols with the $1\sigma$ errorbars represent the magnitudes, while the lines show the best fit to the data. For clarity, the data relative to clump 2.1b and 17.3a have been shifted along the $y$-axis of $-0.5$ and $+1$ magnitudes, respectively.}
    \label{fig:blue_slopes}
\end{figure}

In our sample, we selected eight clumps with robust photometric $\beta$ measurements between $-3.4$ and $-2.8$. We consider only the $\beta$ slopes obtained with at least three flux measurements and without strong nearby contaminants. This subsample, presented in Table~\ref{tab:blue_slopes}, results in the redshift range between approximately 4 and 6 (except for ID 253.3N at redshift $\sim \! 3$) and with a typical $\beta$ uncertainty of $0.4$. The flux measurements and the slopes are shown in Fig.~\ref{fig:blue_slopes}. 
\begin{table*}
\centering
\begin{tabular}{@{}lcccccccc@{}}
\toprule
\toprule
ID & Redshift & $\beta$ slope & \textit{V606} & \textit{I814} &  \textit{Y105} & \textit{J125} & \textit{JH140} & \textit{H160}\\
\midrule
2.1b & 6.15 & $-2.82^{+0.32}_{-0.32}$ &  & & $28.58 \pm 0.08$ & $28.78 \pm 0.12$ & $28.85 \pm 0.12$& $28.89 \pm 0.12$ \\
17.1a & 3.97 & $-3.35^{+0.40}_{-0.40}$ & & $28.76\pm 0.10$& $29.20\pm0.13$& $29.35\pm0.20$ & & \\
17.3a & 3.97 & $-3.08^{+0.35}_{-0.33}$ & & $28.61\pm 0.09$& $28.99\pm0.11$& $29.06\pm0.15$ & & \\
18a & 3.87 & $-3.00^{+0.17}_{-0.16}$ & & $27.90\pm 0.04$& $28.23\pm0.06$& $28.32\pm0.08$ & & \\
70.7N & 5.11 & $-3.10^{+0.56}_{-0.55}$ & & & $29.11\pm 0.13$& $29.34\pm0.20$& $29.44\pm0.21$ & $29.61\pm0.23$ \\
103.1b & 4.12 & $-3.06^{+0.22}_{-0.23}$ & & $28.23\pm 0.05$& $28.50\pm0.07$& $28.78\pm0.12$ & & \\
122 & 6.15 & $-3.06^{+0.36}_{-0.36}$ & & & $28.62\pm 0.08$& $28.75\pm0.12$& $28.93\pm0.14$ & $29.07\pm0.15$ \\
263.3N & 2.93 & $-2.95^{+0.24}_{-0.23}$ &  $28.48\pm 0.09$& $28.69\pm0.08$& $29.13\pm0.13$ & & &\\
\bottomrule
\end{tabular}
\caption{The selected eight clumps with robust extremely blue slopes ($\beta \lesssim -2.7$). The first column contains the ID relative to the clump, as in the catalog by \citetalias{Mestric2022}. The second column displays their redshift, the third the $\beta$ slope measured from photometry, making use of the magnitudes (and $1\sigma$ uncertainties) reported in the others column for the \textit{V606}, \textit{I814}, \textit{Y105}, \textit{J125}, \textit{JH140}, \textit{H160} bands. Each clump has the measurements reported only for the three or four filters that are included in the useful UV rest-frame allowed by its redshift, described by Eqs.~\ref{eq:condition1}-\ref{eq:condition2} and summarized in Table~\ref{tab:bands} and Fig.~\ref{fig:bands2}.}
\label{tab:blue_slopes}
\end{table*}

As shown in the previous section, generally the measured $\beta$ slopes become bluer with increasing redshift and decreasing luminosity, and it has been shown that $\beta$ measurements on photometrically-selected galaxies can likely introduce contamination and biases \citep{Finlator2011, Ceverino2019}. Together with the increasing number of claims of very blue slopes in high-$z$ galaxies, the inspection of the physical properties of their stellar populations became necessary. \citet{Bouwens2010} could reproduce slopes of $\sim \! -3$ with standard \citep{Leiterer1999, Bruzual2003} stellar population models only for very young ($<5$~Myr) star-forming systems and ignoring the nebular continuum emission. The latter is caused by the ionized gas around young stars, and can redden the slopes up to $\Delta \beta \sim 0.5$ \citep{Topping2022}. If this component is included, the slope is not easily reducible below $-2.7$. This suggests that very low metallicity ($Z$) values, or a different IMF, are needed to reproduce more extreme slopes. Some studies \citep[e.g.,][]{Bouwens2010, Maseda2020} found that it is possible to reproduce slopes of about $-3$ with $Z<10^{-2}$ Z$_\odot$, but only for a limited range of ages, between 10 and 30 Myr. They conclude that very low metallicity values can explain part of the extremely blue slopes found, but the limited age range makes it unlikely to be the general explanation. They tested that the obtained slopes are comparable in the most extreme case of a single instantaneous burst with different IMFs and SFHs, like exponential declining or constant star formation.
\cite{Jevrabkova2017} demonstrated that the IMF assumption plays a secondary role in the resulting $\beta$ slopes, but for the youngest ages ($<5$ Myr), where it can account for up to $\Delta\beta = 0.2$. The IMF choice has also a stronger impact when the nebular emission is taken into account. The bluer slopes are related to the presence of the youngest and most massive stars, that would dominate the resulting stellar population if their IMF is top-heavy ($\alpha\sim 1.5$ in $\diff N / \diff m \propto m^{-\alpha}$, where $\alpha=2.35$ in \citealt{Salpeter1955}). The issue of this scenario is that the same stars are incredibly efficient at ionizing the surrounding gas, producing a nebular emission that would make the slope redder than in the young burst scenario with a standard IMF. The contribution from the nebular emission depends on a large number of factors, such as, for instance, the ionization parameter, the metallicity and the geometry. A promising way to decrease it, and thus being able to reproduce more extreme blue slopes, is to consider the case of ionizing radiation that leaks directly into the intergalactic medium (IGM). An escape fraction of ionizing photons into the IGM of 0.3 can easily reproduce the observed blue spectra \citep{Bouwens2010, Zackrisson2013, Chisholm2022}, but this value is considerably larger than the usually assumed $\sim 0.1$, sufficient for galaxies to reionize the Universe. Similarly, \cite{Raiter2010} found that the contribution of the nebular emission strongly affects the slopes, even if also trends with the IMF, SFH, metallicity and age are observed. \cite{Topping2022} explored the possibility that the introduction of binary stars could generate significantly bluer slopes, but could reproduce slopes down to $-3.15$, similar to the $\sim -3.2$ limit reached with single stars, concluding that binaries are not the main responsible for the extremely blue slopes. \\

\begin{figure*}
	\includegraphics[width=0.9\textwidth]{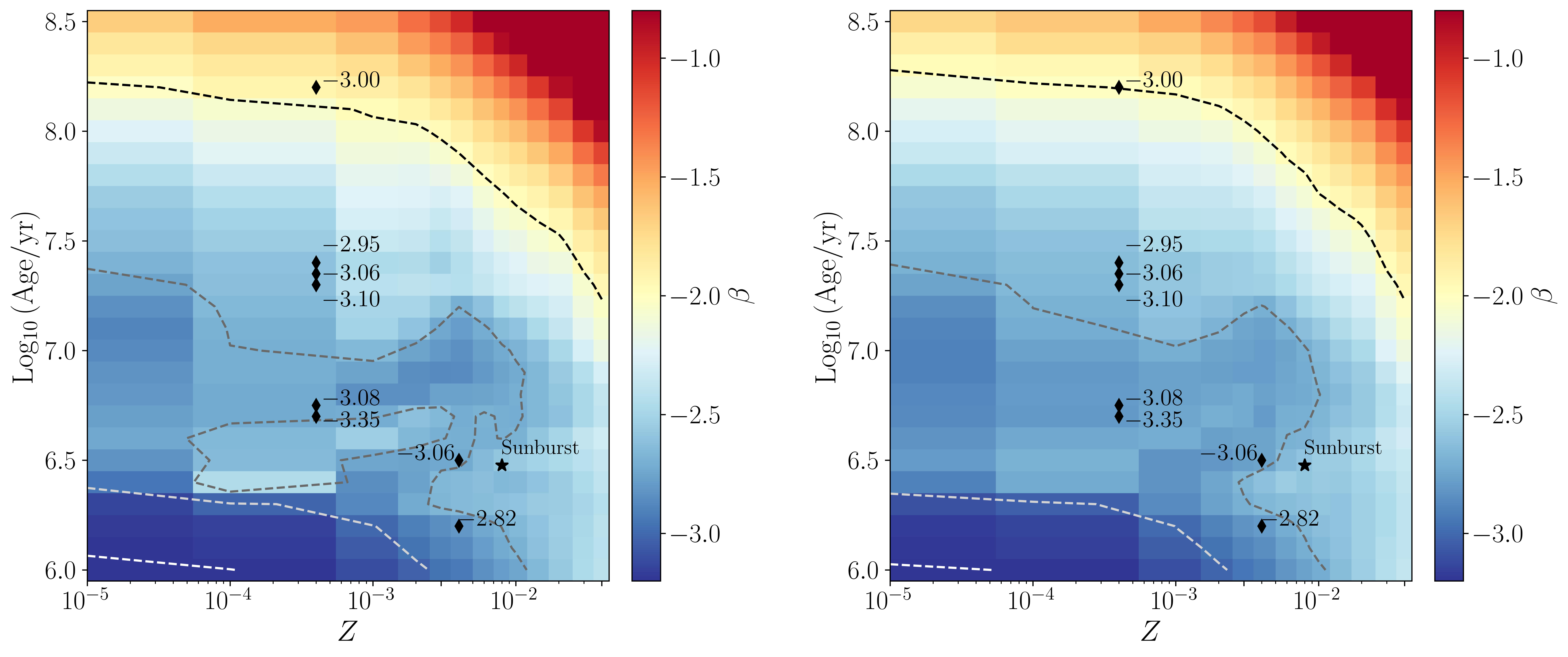}
    \caption{UV continuum $\beta$ slopes (colormap) as a function of the age and metallicity, obtained with the synthetic spectra generated with the \texttt{BPASS} code, not including the nebular emission from the ionized gas around young stars. In this way, it is possible to reproduce the bluest observed slopes, down to $-3.2$, by assuming extremely low metallicity and very young ages (bottom left part of each plot). The $\beta$ values are measured by assuming a pure stellar emission and a Salpeter IMF with 0.5-300 M$_\odot$ mass range (left) and a Chabrier IMF with 1-300 M$_\odot$ mass range with the inclusion of binaries (right). In both panels, the black diamonds represent our eight extremely blue clumps, with their $\beta$ slopes measured from photometry (see Table~\ref{tab:blue_slopes}). The age and metallicity values are the best-fit measured by \citetalias{Mestric2022} through SED modelling, with a typical uncertainty on the age of $+1.7$ and $-0.7$ dex. The black star represents the stellar cluster 5.1 of the Sunburst galaxy, at $z=2.37$ \citep{Dahle2016}. We measure a photometric $\beta$ slope of $-2.41 \pm 0.01$, exploiting the F555W, F606W, and F814W \textit{HST} filters. The dashed lines, from black to white, represent the [$-2$, $-2.7$, $-3$, $-3.2$] contours, respectively.}
    \label{fig:bpass}
\end{figure*}

We analyze the physical properties of our sample of extremely blue clumps making use of the publicly available Binary Population and Spectral Synthesis code \citep[\texttt{BPASS v2.3},][]{Byrne2022, Eldridge2017} through its Python version Hoki \citep{Stevance2020}, which implements binary stellar evolution models and synthetic stellar populations to investigate the properties of the integrated light emitted from physically motivated distant stellar populations. We measured the $\beta$ slopes of distant galaxies from their synthetic spectra with the same procedure described in Sect.~\ref{sect:spectroscopic}. The different spectra are obtained by varying the main physical parameters that impact the $\beta$ values: the metallicity, the age, the presence of binaries, and the IMF. We assumed four different IMFs: 1) a Salpeter IMF \citep{Salpeter1955} with 0.5-100 M$_\odot$ mass range; 2) a Salpeter IMF with 0.5-300 M$_\odot$ mass range; 3) a Chabrier IMF with 1-100 M$_\odot$ mass range; and 4) a Chabrier IMF with 1-300 M$_\odot$ mass range, and measured the $\beta$ slopes over a grid of metallicity ($Z=10^{-5}$, $10^{-4}$, 0.001, 0.002, 0.003, 0.004, 0.005, 0.006, 0.008, 0.010, 0.014, 0.020, 0.030, 0.040) and age (from $\log_{10}(\mathrm{age/yr})=6.0$ to 8.5, with 0.1 steps) values. We repeat each configuration including the presence of binaries. In order to reproduce the bluer slopes, we focus only on pure stellar emission models, not considering the nebular emission. We obtain similar trends for all the models, and we show the results for two of them, with a Salpeter IMF with 0.5-300 M$_\odot$ mass range, a Chabrier IMF with 1-300 M$_\odot$ mass range and including binaries, in Fig.~\ref{fig:bpass}. The minimum $\beta$ value we could reach is approximately $-3.2$ for all the models, with the absolute minimum $-3.22$ value obtained for the model showed on the left of Fig.~\ref{fig:bpass}. The introduction of binary systems makes on average the slopes bluer of $\Delta \beta \sim 0.08$ for all the IMFs, for all the age and metallicities. The extension of the IMF mass range from 1-100 M$_\odot$ to 1-300 M$_\odot$ makes them bluer of $\Delta \beta \sim 0.02$. For all the models, this mean value affects uniformly all the ages and metallicities, but for $\log_{10}(\mathrm{age/yr}) \lesssim 6.5$, where $\Delta \beta$ has a mean value of 0.2, with a peak of 0.4. In both the panels of Fig.~\ref{fig:bpass}, the black diamonds represent our eight extremely blue clumps according to the best-fit results obtained by \citetalias{Mestric2022} through SED modelling, with a mean uncertainty on the age of $+1.7$ and $-0.7$ dex. The measurements of the age suffer from large uncertainties because of the young ages of these clumps, and we would need future spatially resolved spectroscopic observations to better constraint these quantities and directly compare models and observations. In the same plot, we mark, with a black star, the location of the stellar cluster 5.1 hosted by the Sunburst galaxy, at $z=2.37$ \citep{Dahle2016, Chisholm2019Chisholm}. This stellar cluster, with 12 multiple images, presents a multi-peaked Ly$\alpha$ emission that is consistent with an optically thin medium and Lyman continuum (LyC) leakage along the line of sight \citep{Rivera-Thorsen2017}. Additional studies revealed that this stellar cluster is younger than 3 Myr and presents a stellar metallicity of 0.4 Z$_\odot$, with a physical size of $\approx 10$ pc and a stellar mass of $\approx 10^7$ M$_\odot$ \citep{Vanzella2022, Mestric2023}. The observed LyC leakage is consistent with low nebular emission, and it makes it possible to compare the $\beta$ slope of this system with those measured from our \texttt{BPASS} models, where the nebular emission contribution is not included. Sunburst benefits from a comprehensive collection of photometric and spectroscopic data \citep{Mestric2023}. To be fully consistent with the presented results, we measure the photometric $\beta$ slope with the same approach described here, exploiting the F555W, F606W, and F814W \textit{HST} filters. The UV continuum slope we measure is $\beta = -2.41 \pm 0.01$.

\subsection{Caveats}
\label{subs:caveats}
The measurement of the UV continuum $\beta$ slope from photometry over a wide range of redshift values suffers from several well-known biases. In the study by \cite{Bouwens2014}, a comprehensive examination of potential systematic uncertainties affecting the derived $\beta$ slopes reveals a multitude of small factors. These factors include uncertainties in the effective PSFs of the HST observations, errors in accurately registering the observations with each other, the derivation of PSF kernels to ensure consistency across multiple bands, uncertainties in the HST zeropoints, the influence of light emitted by neighboring sources, and potential systematic errors in background subtraction. When the different images are PSF matched with the procedure we followed, \cite{Bouwens2014} estimate a total systematic uncertainty of about $3\%$ in the measured colors. \\
One of the most relevant systematics effects is the ``blue bias'' \citep[e.g.,][]{Dunlop2012, Rogers2013, Bouwens2014, Jiang2020, Bhatawdekar2021, Cullen2022}, that makes the faintest galaxies to have bluer slopes. This effect is due to the selection of candidate high-$z$ galaxies by using filters close to the Lyman-$\alpha$ emission line, that enhances the flux in the short wavelength part of the spectrum and makes the slope bluer. In our study, the effect of this bias is absent, thanks to the spectroscopic confirmation of all the clumps. Furthermore, the selection of the exploitable filters that are included in the rest-frame UV, is properly designed to avoid Lyman-$\alpha$ contaminations (see Eqn.~\ref{eq:condition1} and \citealt{Calzetti1994}). Despite this, our sample shows a correlation of the measured $\beta$ slopes with the UV magnitude: it is usually interpreted as a change in the metallicity and in the dust extinction, but a contribution related to this bias cannot be excluded. Another observed systematic effect, of the order of $\Delta \beta \sim 0.2-0.3$, is related to the wavelength range in which the slopes are measured. In our study, and consistently with broad band measurements, we exploit the entire UV wavelength range, but it was shown that $\beta$ values measured between $1300-1800\, \si{\angstrom}$ and $1800-2200\, \si{\angstrom}$ can be slightly different \citep[e.g.,][]{Raiter2010, Chisholm2022}. This effect may become particularly relevant when comparing the photometric $\beta$ slopes with the spectroscopic ones or, depending on the redshift of the source, when the used photometric filters do not cover the entire UV range. This effect is stronger in the case of $\beta$ slopes measured from a small number of available fluxes, as it can be the case of our fits with two or three magnitude measurements \citep[e.g.,][]{Jiang2020, Mondal2023}. We studied in detail the effects of the number of filters used in the fit, giving particular attention to the 2-magnitudes fits. Even if they are commonly employed in this kind of studies, given that no more of four HST broad filters can be simultaneously included in the relatively narrow restframe UV wavelength range, we observed significant trends, in particular regarding the uncertainties and the extreme $\beta$ values. The 2-magnitude fits, even if they have been derived for the lowest redshift and brightest clumps, have the largest uncertainties. The tails of the distribution of the measured $\beta$ slopes are strongly dominated by those clumps (see also Fig.~\ref{fig:beta_hist}). For this reason, we decided to exclude 2-magnitudes fits from some parts of the analysis, as described in the previous sections, and we checked whether all the results and correlations would importantly change by including or excluding them. 

\section{Summary and conclusions}
We measured the UV-continuum $\beta$ slopes of a sample of 166 individual star-forming clumps, belonging to 67 galaxies strongly lensed by the cluster of galaxies MACS~J0416.1$-$2403, making use of PSF-matched HST photometry for the entire sample, joint with deep MUSE spectroscopic observations for 100 clumps of the sample. We accurately analyzed and discussed the possible presence of biases and systematic uncertainties on the $\beta$ measurements. The first aim of this study is to compare our novel measurements for individual clumps with those for integrated galaxies, in order to investigate possible physical differences between these regions and their hosts. We pursue it by measuring the value of the UV-continuum $\beta$ slope ($f_\lambda \propto \lambda^\beta$), which depends on different key physical parameters, such as the age, metallicity, dust extinction, IMF and SFH. As is common in analogous studies referred to high-$z$ galaxies, we investigate the trends of the $\beta$ values as a function of the redshift and the UV luminosity. Our main conclusions can be summarized as follows. 

\begin{itemize}
    \item The $\beta$ slope distribution of our sample of individual clumps shows a median value of $\sim -2.4$, with a standard deviation of $\sim 0.8$. This value is bluer than the $\sim -2$ value measured in literature for integrated galaxies in the same redshift range. The bluer median value for individual clumps confirms that they are sites of star formation, populated by young, massive OB stars, whose spectrum strongly emits in the UV, but can also point out different features between the host galaxy and their clumps. In fact, the bluer median slope can suggest a dishomogeneous dust distribution in the galaxy, and that the dust at the location of clumps is lower than the average extinction of the galaxy. Also, it can be explained by assuming a different SFH or SFR recipe for the clumps. 

    \item The measured $\beta$ values show a trend with the absolute magnitude in the restframe UV, $M_{UV}$, consistent with the relation expressed by \citet{Bouwens2014}. They show a scatter of $\Delta \beta \sim$ 0.26, 0.15, and 0.35, in the $z \sim 4$, $z \sim 5$ and $z \sim 6$ bin, respectively. The reference relation was measured for high-$z$ galaxies, with $M_{UV}$ ranging from $-22$ to $-16$, while our sample of clumps covers $M_{UV}$ from $-18$ to $-12$. 
    This implies that this relation can be extended to much fainter magnitudes, and that clumps follow the same relation as their host galaxies (see Fig.~\ref{fig:beta-Muv}). 

    \item We observe a weak trend of the $\beta$ values with the redshift, as observed for integrated galaxies. We fit our entire sample and measure the relation $\beta = (-0.57 \pm 0.05) z + (-0.47 \pm 0.12)$, which is steeper than that measured by \cite{Bouwens2014}. But, they are obtained in different redshift bins and, limiting our fit to the $z \gtrsim 3.5$ clumps, we obtain a much flatter slope of $(-0.26 \pm 0.09)$, consistent with the results for integrated galaxies.

    \item In the $\beta$ slopes distribution of our sample of individual clumps, several objects populate the tails of this distribution: 4 objects have a very red ($\beta > -1$) and 30 a very blue ($\beta < -2.7$) slope. We focused on eight objects with very blue robust $\beta$ slope, obtained by fitting magnitude measurements in at least three different filters. We used the code \texttt{BPASS}, that simulates stellar populations and follows their evolution, to generate synthetic spectra of galaxies with different metallicities, SFHs, IMFs, and the possible presence of binaries. We were able to reproduce slopes down to $\beta \sim \-3.2$, by assuming low metallicity ($Z \lesssim 10^{-3}$), young ($\log{\mathrm{(age/yr)}} \lesssim 7$) and dust-poor regions, considering the absence of the nebular emission, whose presence would not allow us to reach so blue slopes, reddening them typically by $\Delta \beta \sim 0.5$.

\end{itemize}

Even if based on some of the deepest and best observations currently available for lensed fields, this study could be improved in several ways with additional data from current and new facilities. First of all, the sample of individual clumps can be enhanced by including both other fields lensed by cluster of galaxies, and non-lensed galaxies. Moreover, it will allow us to improve the measured distributions by adding catalogs with similar depths and redshift ranges. Then, the redshift and the magnitude ranges can be enlarged thanks to infrared coverage of JWST, that will be able to measure $\beta$ slopes with approximately $\sigma_\beta \sim 0.2$ uncertainty for $M_{UV}<-20$ at $z>8$. Concerning the extremely blue slopes, an extension of the synthetic models explored and a broad-wavelength spectroscopic follow up with ground (e.g., VLT/X-Shooter) or space (i.e., JWST/NIRSpec) instruments of the bluest and brightest clumps represent essential steps in the study of the first galaxies and of the epoch of reionization. A first robust confirmation of galaxies, or isolated clumps, with uncommonly low metallicity or dust extinction values can reshape and deepen our comprehension on how galaxies were born, how they evolve, as well as the fate of their star-forming clumps. 

\section*{Acknowledgements}

We acknowledge financial contributions from PRIN-MIUR 2017WSCC32 and 2020SKSTHZ. 
EV acknowledges support from the INAF GO Grant 2022 “The revolution is around the corner: JWST will probe globular cluster precursors and Population III stellar clusters at cosmic dawn”. 
MC acknowledges support from the INAF Minigrant ``Reionization and fundamental cosmology with high-redshift galaxies". 
FC acknowledges support from PRIN INAF 1.05.01.85.01 and from grant PRIN MIUR 20173ML3WW$\_$001. 
EI acknowledges funding from the Netherlands Research School for Astronomy (NOVA).

\section*{Data Availability}
The data underlying this article will be shared on reasonable request to the corresponding author.


\bibliographystyle{mnras}
\bibliography{biblio} 


\appendix

\section{Location of the mock clumps in the simulation}
\label{app:extra_sim}
\begin{figure*}
	\includegraphics[width=0.7\textwidth]{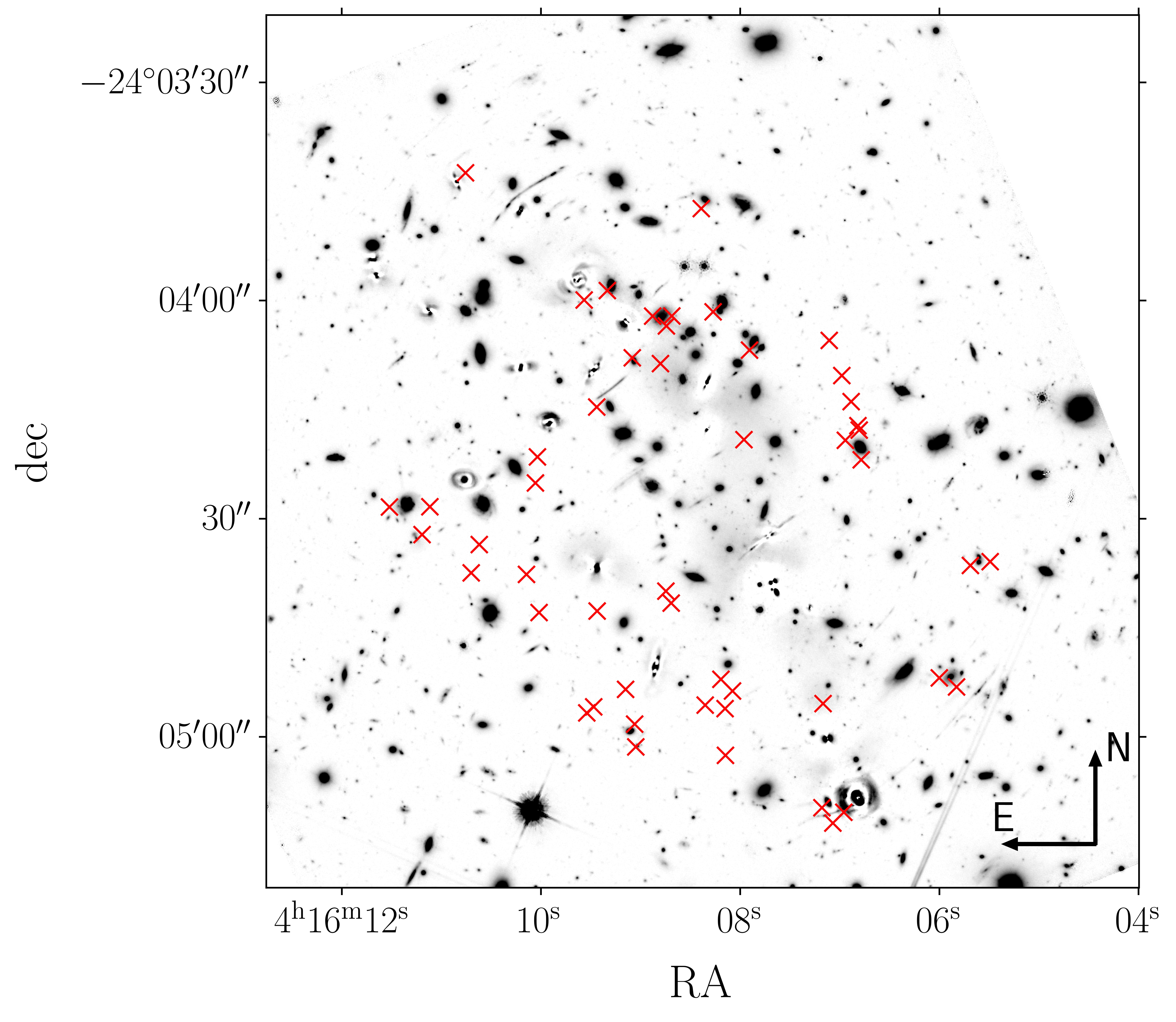}
    \caption{Location of the 50 mock clumps (red crosses) superimposed to the F105W image of the lens cluster MACS~0416. We locate them in positions similar to those of the real clumps. Hence, we choose the outer and inner regions of the cluster, to see the possible residuals from the intracluster light removal, in isolated positions, and angularly close to a bright object, to quantify the contribution of the contamination of foreground galaxies. In this case, we put the mock clump at the same angular distance to the contaminant as that of the real clump, but in an opposite direction, to avoid the real clumps to contaminate the simulation.}
    \label{fig:sim_loc}
\end{figure*}

In order to find the best combination of parameters to measure the $\beta$ slope of the clumps in our sample, we test the possible presence of systematics on a sample of 50 mock clumps, placed in different locations around the lens cluster MACS~J0416, as shown in Fig.~\ref{fig:sim_loc}. They are not randomly distributed, but their positions are accurately chosen to investigate where the $\beta$ measurements might be biased. In particular, we check the impact of the contribution of the ICL and of the presence on angularly close foreground bright objects. Then, we look for the best combination of A-PHOT parameters that is able to minimize the difference between the injected $\beta$ slope and that measured. In particular, we test different apertures, from a diameter of $0.2 \arcsec$ to $0.54\arcsec$, to switch on and off the A-PHOT local background estimation, and to manually fit the light and subtract the possible foreground contaminant and the background level with GALFIT.

\section{Comparison with {\tt ASTRODEEP}}
\begin{figure*}
	\includegraphics[width=\columnwidth]{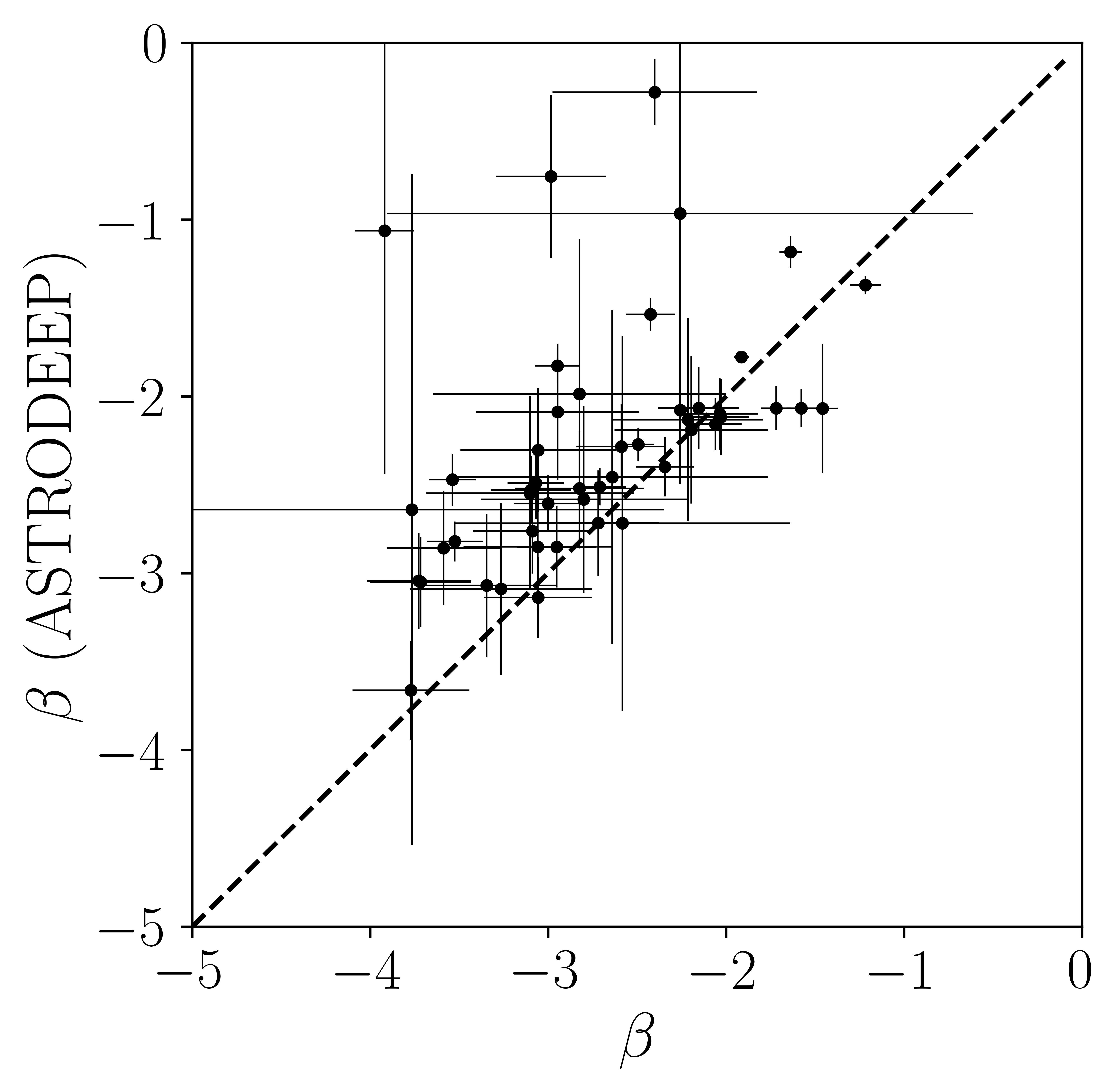}
    \caption{Comparison between the $\beta$ slopes measured with the magnitudes measured from the {\tt ASTRODEEP} collaboration ($y$-axis) and in this work ($x$-axis). {\tt ASTRODEEP} slopes are systematically redder (median $\Delta \beta \sim 0.24$), and thus lay above the 1:1 relation (dashed line), but they are referred to integrated galaxies extracted from larger apertures, while we isolated individual hosted clumps. }
    \label{fig:comparison_AD}
\end{figure*}

We apply our pipeline to estimate the $\beta$ slopes of our sample but making use of the photometric measurements by the {\tt ASTRODEEP} collaboration \citep{Merlin2016a, Castellano2016}, that detected and characterized the objects in the MACS~J0416 field. \citetalias{Mestric2022} cross-matched our sample with the {\tt ASTRODEEP} catalog, finding 48 objects in common. The comparison with the $\beta$ slopes obtained with our photometric measurements is shown in Fig.~\ref{fig:comparison_AD}.

\section{Comparison between photometric and spectroscopic $\beta$ slopes}
\begin{figure*}
	\includegraphics[width=0.8\textwidth]{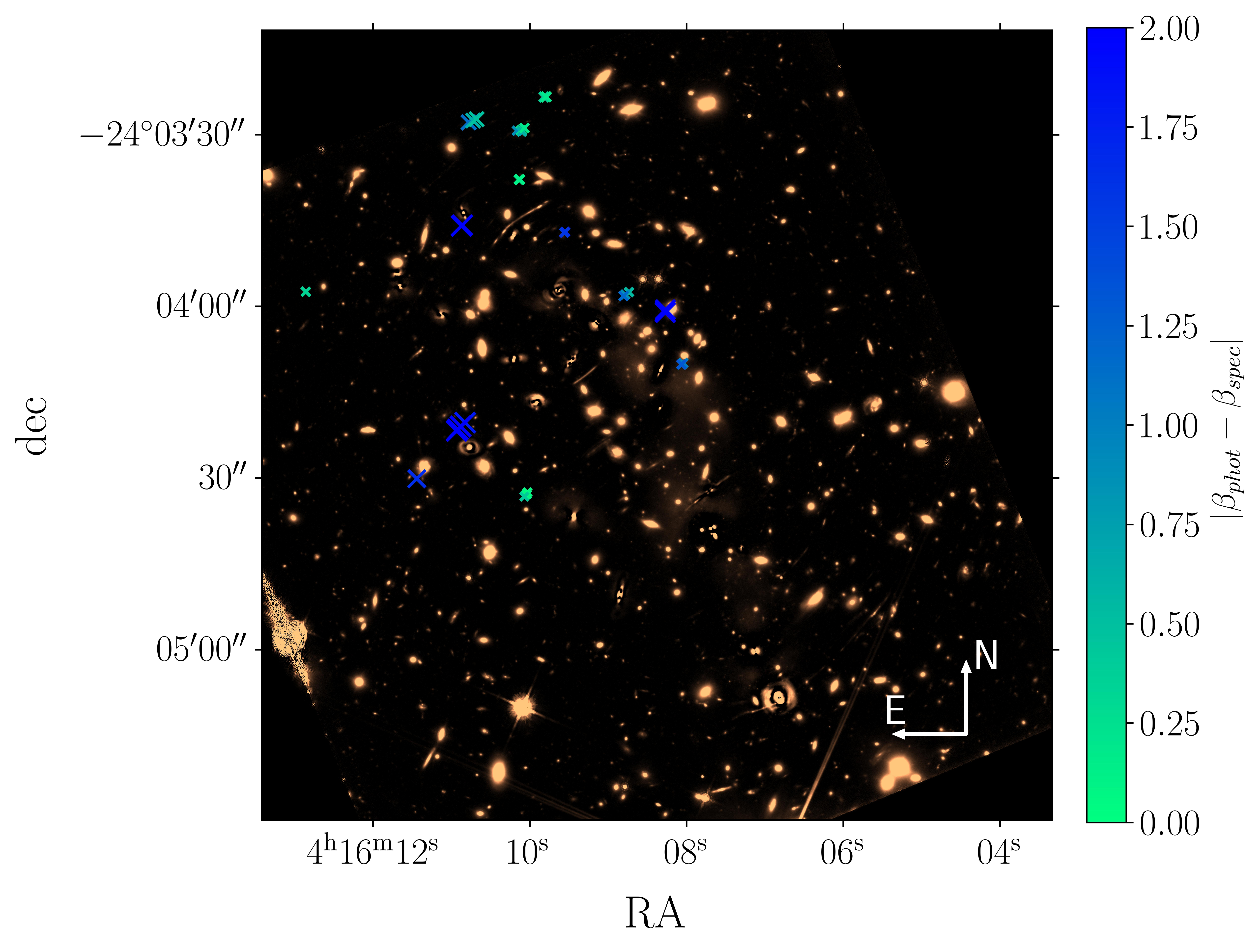}
    \caption{Location of the 37 clumps whose spectra have $S/N\gtrsim 2$, on the \textit{I814} image. The marker size depends on the redshift, increasing from 1.99 to 3.29. They are color-coded following the difference between the spectroscopic ($\beta_{spec}$) and photometric ($\beta_{phot}$) $\beta$ slopes. The most discrepant clumps appear angularly close to bright contaminants, which are subtracted in the photometric images but not in the MUSE datacube. }
    \label{fig:specVSphot_position}
\end{figure*}

We compare the $\beta$ slopes measured with photometry and with spectroscopy for a subsample of 37 clumps, whose spectrum has $S/N\gtrsim 2$. We observe that spectroscopic ones result to be systematically redder. We explain this result by considering the contamination from some red foreground objects, the BCG, and the ICL, which are not subtracted in the MUSE datacube, unlike the HST images, and the larger aperture (photometry extracted from a 0.27\arcsec diameter aperture, spectra extracted from 0.4\arcsec-diameter ones) that enhance this effect. We show, in Fig.~\ref{fig:specVSphot_position}, that indeed, the most discordant slopes are measured for clumps angularly close to a foreground contaminant or located in the central regions of the cluster of galaxies.

\bsp	
\label{lastpage}
\end{document}